\documentclass[journal]{IEEEtran}

\usepackage{graphicx}      

\usepackage{color}
\usepackage{cite}

\usepackage{amsmath}

\usepackage{amssymb}  

\usepackage{algpseudocode} 

\newcommand{\bbaru}{\boldsymbol{\bar{u}}}
\newcommand{\bbarx}{\boldsymbol{\bar{x}}}

\newcommand{\bbarg}{\boldsymbol{\bar{\gamma}}}
\newcommand{\bu}{\boldsymbol{u}}

\newcommand{\bbarv}{\boldsymbol{\bar{v}}}

\newcommand{\bbare}{\boldsymbol{\bar{\eta}}}
\newcommand{\be}{\boldsymbol{\eta}}

\newcommand{\bz}{\boldsymbol{z}}

\newcommand{\bx}{\boldsymbol{x}}
\newcommand{\by}{\boldsymbol{y}}

\newcommand{\bd}{\boldsymbol{\delta}}

\newcommand{\xig}{\xi_{\mathcal{G}}}
\newcommand{\etag}{\eta_{\mathcal{G}}}
\newcommand{\diag}{\operatorname{diag}}
\newcommand{\one}{\mathbf{1}}

\newtheorem{prob}{Problem}
\newtheorem{defn}{Definition}
\newtheorem{assum}{Assumption}
\newtheorem{thm}{Theorem}
\newtheorem{lem}{Lemma}
\newtheorem{prop}{Proposition}
\newtheorem{rem}{Remark}
\newcommand{\interior}{\operatorname{int}}

\usepackage{lipsum}                     
\usepackage{xargs}                      
\usepackage[pdftex,dvipsnames]{xcolor}  
\usepackage[colorinlistoftodos,prependcaption,textsize=tiny]{todonotes}
\newcommandx{\unsure}[2][1=]{\todo[linecolor=red,backgroundcolor=red!25,bordercolor=red,#1]{#2}}
\newcommandx{\change}[2][1=]{\todo[linecolor=blue,backgroundcolor=blue!25,bordercolor=blue,#1]{#2}}
\newcommandx{\info}[2][1=]{\todo[linecolor=OliveGreen,backgroundcolor=OliveGreen!25,bordercolor=OliveGreen,#1]{#2}}
\newcommandx{\improvement}[2][1=]{\todo[linecolor=Plum,backgroundcolor=Plum!25,bordercolor=Plum,#1]{#2}}
\newcommandx{\thiswillnotshow}[2][1=]{\todo[disable,#1]{#2}}

\allowdisplaybreaks

\begin{document}
	\title{An optimization-based cooperative path-following framework for multiple robotic vehicles}
	\author{Andrea Alessandretti and 
		A. Pedro Aguiar  
		\thanks{This work was supported in part by projects IMPROVE-POCI-01-0145-FEDER-031823 and HARMONY-POCI-01-0145-FEDER-031411 - funded by FEDER funds through COMPETE2020-POCI and by national funds (PIDDAC). A. Pedro Aguiar received a Sabbatical grant from FCT (Funda\c{c}\~ao para a Ci\^encia e a Tecnologia).}
		\thanks{Andrea Alessandretti and A. Pedro Aguiar are with the Faculty of Engineering, University of Porto, Porto, Portugal. E-mails: \{andrea.alessandretti,pedro.aguiar\}@fe.up.pt}
	}
	\maketitle
	
	\begin{abstract}This paper addresses the design of an optimization-based cooperative path-following control law for multiple robotic vehicles that optimally balances the transient trade-off between coordination and path-following errors. To this end, we formulate a more general multi-agent framework where each agent is associated with \emph{(i)} a continuous-time dynamical model, which governs the evolution of its state, and \emph{(ii)} an output equation that is a function of both the state of the agent and a coordination vector. According to a given network topology, each agent can access its state and coordination vector, as well as the coordination vectors of the neighboring agents. In this setup, the goal  is to design a distributed control law that steers the output signals to the origin, while simultaneously driving the coordination vectors of the agents of the network to consensus. To solve this, we propose a model predictive control scheme that builds on a pre-existing auxiliary consensus control law to design a performance index that combines the output regulation objective with the consensus objective. Convergence guarantees under which one can solve this coordinated output regulation problem are provided. Numerical simulations display the effectiveness of the proposed scheme applied to a cooperative path following control problem of a network of 3D nonholonomic robotic vehicles.
	\end{abstract}

	\IEEEpeerreviewmaketitle

	\section{Introduction}
	Important applications using multiple autonomous robotic vehicles that arise in a number of scenarios including searching, surveying, and exploration, motivate the formulation of a particular class of motion control tasks that can be described as cooperative path-following (CPF) problems.
	
	In simple terms, given a finite set of vehicles and conveniently desired geometric paths (one per vehicle), and an underlying communication topology that establishes the communication between them, the CPF problem consists of designing the motion control algorithms for each vehicle to drive and maintain them in their respective desired paths with a common speed profile and holding a specified formation pattern.  
	Different solutions to this problem and similar problems can be found in the literature, see e.g.,
\cite{aguiar2007coordinated,aguiar2008coordinated,beard2001a,Borhaug11, cichella2016safe,Egerstedt01,ghabcheloo2009coordinated,Ihle07, skjetne2004robust,xiang2012synchronized}, and the references therein. 
	
	%
	A particular interesting strategy to solve the problem consists of decoupling it into \emph{(i)} a path following goal, where the objective is to derive control laws to drive each vehicle to its path at the reference speed profile, and \emph{(ii)} a multiple vehicle coordination task, where the objective is to adjust the speed of each vehicle so as to achieve the desired formation pattern by running a consensus like algorithm that should take explicitly into account the topology of the inter-vehicle communications network.
	The works in \cite{aguiar2007coordinated,ghabcheloo2009coordinated} offer a theoretical overview of the subject and introduce techniques to solve this problem. In those works, a key idea is to parameterize by a single variable each desired geometric path with the restriction that when all the vehicles are following the respective desired paths in the required formation pattern, the parameterization variables have the same value, that is, the path variables are all synchronized. Then, control laws are independently designed for path-following and consensus.

	In this paper, contrary to the works mentioned above, we propose an optimization-based CPF design methodology that optimally balances in an integrated form the transient trade-off between coordination and path-following errors. 
	To this end, we consider a more general framework that captures the case of a multi-agent system where each agent is associated with a continuous dynamical model that includes an output equation that is a function of the state of the agent and a coordination vector. For a given network topology, each agent can access its state and coordination vector, as well as the coordination vectors of the neighboring agents. The goal in this setup is to design a distributed control law that steers the output signals to the origin, while simultaneously driving the coordination vectors of the agents of the network to consensus. 
	To solve this, we propose a model predictive control scheme that builds on a pre-existing auxiliary consensus control law to design a performance index that combines the output regulation objective with the consensus objective. It is important to stress that 
	this coordinated output regulation problem, which is formally introduced in Section \ref{sec:problem_definition}, stands at the intersection of the consensus problem in multi-agent systems (MAS) and the output stabilization problem.

	The consensus problem addresses the design of distributed control laws for the agents of a MAS to be able to steer the state of each agent to a common value. The control law is distributed, in the sense that for a given network topology each agent can only use measurements of its state and the state of the neighboring agents. The high relevance of such multi-agent system coordination problem is well motivated by  
the existence of an extensive literature on the topic, and we refer to, e.g., \cite{bertsekas1989parallel,consensus02,Olfati-Saber2007,Mesbahi2010a} for an introduction to it.
	
	The Model Predictive Control (MPC) solves the stabilization problem by selecting among the future feasible input signals, the one that minimizes a given (and suitable) performance index defined in a finite horizon, applying the first part of the optimal input signal, and iterating the process.
	 For MPC solutions to the trajectory tracking and path-following problem, we refer the works in \cite{Faulwasser2015,Gu2006,Alessandretti2013}. For a general survey on MPC, the works \cite{Mayne2000,rawlings2009model,Mayne2014}, and \cite{Scattolini2009a} for a classification of different MPC architectures in MAS.

	One of the main challenges in the design of MPC schemes for MAS stems from the fact that every agent can only predict its own future state and input trajectory without the knowledge of the future behavior of the neighboring systems. To address this problem, in \cite{Dunbar2006,Wang2010,Farina2012a}, the controllers employ a specific constraint, often called consistency constraint, devoted to guaranteeing that the optimal prediction of each agent stays close to the one obtained at the previous time instant, which was exchanged among neighboring agents. This information is then used in the analysis to obtain closed-loop guarantees. For the case of systems affected by noise, the work in \cite{Wang2010} provides guarantees of convergence to the minimal disturbance invariant set. For the case of single-/double-integrator systems, the method proposed in \cite{Ferrari-Trecate2009} uses a performance index that prompts a contraction of the states of the network and therefore, building on \cite{Moreau2005}, guarantees convergence to consensus.

In this work, for a specific class of systems, we propose a different MPC framework that does not require sharing the prediction of the state with the other neighboring systems and thus does not employ consistency constraints, but it only shares the coordination vectors that indirectly define when the systems are in coordination. Because of this, the knowledge of the dynamical model of the neighboring agents, their state, or prediction of the state, are not required, and only the value of their coordination vectors is shared.  In addition, building on a pre-existing auxiliary consensus control law, only the sharing of the current value of the coordination vector, and not its prediction, is required. 

The propose optimization-based control law is able to drive the output signals of the agents in a MAS to the origin, where such output is a function of a coordination vector that we wish to drive to consensus. In this approach, one can consider heterogeneous agents, where each agent may have a possible different dynamical model, and we provide convergence guarantees under which it is possible to solve the proposed MAS coordinated output regulation problem. We focus on distributed non-iterative MPC schemes, i.e., the information is only exchanged among neighboring agents, and the communication is only performed once per every solution of the MPC optimization problem. This is in contrast to iterative methods, where the MPC controllers need to communicate multiple times to solve the local MPC problem. Furthermore, the proposed scheme falls in the category of non-cooperative MPC schemes, where each agent minimizes an independent performance index. In the single-agent case, the proposed scheme also provides a solution to the trajectory tracking and path following problem. 

A subset of the results reported here was presented in \cite{alessandretti2017distributed}. In
this paper, we focus more on the CPF framework problem, relax the assumptions on the model formulation and design procedure, and provide an extended proof of the results.
	
	The remainder of the paper is organized as follows. Section~\ref{sec:problem_definition} starts with the description of the motivated CPF example for a class of 3D nonholonomic robotic vehicles and ends with the formulation of the more general coordinated output regulation problem for multi-agent systems.
	Next, Section~\ref{sec:main} proposes the optimization-based control framework with convergence guarantees. For the sake of clarity, all the proofs are reported in the Appendix. In Section~\ref{sec:numerical}, we particularize the framework for the motivating example and present numerical simulations that show the effectiveness and performance of the proposed scheme.  Conclusions 
	are presented in Section \ref{sec:Conclusions}.
	
	{\bf Notation and definitions:} The term $\mathcal{C}(a,b)$ and $\mathcal{PC}(a,b)$ denotes the space of continuous and piecewise continuous trajectories, respectively, defined over $[a,b]$ or $[a,+\infty)$ for the case where $b=+\infty$. For a generic time-dependent trajectory, we use the bold notation to refer to the whole trajectory and the normal notation to refer to the trajectory evaluated at a specific point in time, e.g., for a continuous-time trajectory $x:\mathbb{R}_{\geq t_0}\to \mathbb{R}^n$ we write $\bx\in\mathcal{C}(t_0,+\infty)$ and $x(t)\in\mathbb{R}^n$, with $t\geq t_0$. For a given $n\in\mathbb{N}$, $SE(n)$ denotes the Cartesian product of $\mathbb{R}^n$ with the group $SO(n)$ of $n \times n$ rotation matrices and $se(n)$ denotes the Cartesian product of $\mathbb{R}^n$ with the space  $so(n)$ of $n \times n$ skew-symmetric matrices. For a generic time $t\geq t_0$, the superscript $^{\star t}$ is used to denote all the trajectories of a given signal associated with the optimal predictions of $\mathcal{P}(t)$. The terms $I_{n}$ and $\one_n$ denote the $n$-by-$n$ identity matrix and the $n$-by-$1$ column vector with all components equal to $1$. For a generic set $\mathcal{S}$ the term $\interior(\mathcal{S})$ denotes the interior of $\mathcal{S}$. The term $\mathcal{B}(r)$ denotes the closed ball set of radius $r\geq0$, i.e.,  $\mathcal{B}(r):=\{x:\|x\|\leq r\}$, where for a generic vector $v$ and matrix $M$ the terms $\|v\|=\sqrt{v'v}$ and $\|M\|=\max_{v\neq0}(\|Av\|/\|v\|)$ denote the vector norm and the associated induced matrix norm, respectively. In addition, for a square matrix $A$, we define $\|v\|^2_A:=v'Av$. 
	For a generic matrix $A$, $[A]_i$ denotes its $i$-th row vector. Given a set $\mathcal{T}:=~\{t_0, t_1,\dots \}\subset\mathbb{R}$ of time instants, we denote with $\lfloor t \rfloor $ the maximum sampling instant $t_k\in\mathcal{T}$ smaller than or equal to $t$, i.e., 
	$
	\lfloor t \rfloor  = \max_{k\in\mathbb{N}_{\geq0}} \{t_k~\in~\mathcal{T}: t_k~\leq~t\}.
	$
	We also use the simplified notation of omitting the time dependence and in particular, for the $\tau$ dependent signals $\bbarx^b_a$, $\bbaru^b_a$, $\bbarg^b_a$  $\bbarv^b_a$, a function $l(\cdot)$ evaluated as $l(\tau,\bar{x}^b_a(\tau),\bar{u}^b_a(\tau),\bar{\gamma}^b_a(\tau),\bar{v}^b_a(\tau))$ can be denoted by $l(\tau,\bar{x}^b_a,\bar{u}^b_a,\bar{\gamma}^b_a,\bar{v}^b_a)$ or $\bar{l}^b_a(\tau)$, whenever clear from the context.

	\vspace{-0.2cm}
	\section{Motivating Example and Problem Statement} \label{sec:problem_definition}
	
In this section, as a motivating example, we formulate a cooperative path-following control problem for a class of 3D nonholonomic robotic vehicles that fits the more general problem that will be described afterwards. 
\vspace{-0.2cm}
	\subsection {Cooperative path-following example} \label{sec: CPF robot}
	Let $\{I\}$ be an inertial coordinate frame and $\{B\}^{[i]}$  a body coordinate frame attached to the generic vehicle $i$. The pair $(p^{[i]}(t),R^{[i]}(t)) \in SE(3)$ denotes the configuration of the vehicle, position and orientation, where  $R^{[i]}(t)$ is the rotation matrix from  body to inertial coordinates. Now, let $(v^{[i]}(t),\Omega(\omega^{[i]}(t)))~\in~se(3)$ 
	be the twist that defines the velocity of the vehicle, linear and angular, where the matrix $\Omega(\omega^{[i]}(t))$ is the skew-symmetric matrix associated with the angular velocity $\omega^{[i]}(t):= \begin{bmatrix} \omega_1^{[i]}(t)& \omega_2^{[i]}(t)& \omega_3^{[i]}(t)\end{bmatrix}'
	$, defined as
	$
	\Omega(\omega^{[i]}) := {\footnotesize
	\begin{pmatrix}
	0  &  -\omega_3^{[i]} & \omega_2^{[i]} \\
	\omega_3^{[i]}  &  0 & -\omega_1^{[i]}\\
	-\omega_2^{[i]} & \omega_1^{[i]} & 0
	\end{pmatrix}\nonumber.}
	$
	The kinematic model of the body frame satisfies
		\vspace{-1mm}\begin{align}
		\dot{p}^{[i]}(t) &= R^{[i]}(t)v^{[i]}(t),& \dot{R}^{[i]}(t) = R^{[i]}(t)\Omega(\omega^{[i]}(t)).\label{eq:model}
		\end{align} 
		In this example, we consider the nonholonomic case with the only actuation in the forward linear velocity, that is,  
		$v^{[i]}(t)~=~\begin{bmatrix}~v^{[i]}_{1}(t)& 0& 0\end{bmatrix}'$, and in the two components of the angular velocity associated with the pitch and yaw motion, that is, $\omega_1^{[i]}(t)=0$. Therefore, the control input is given as $
		u^{[i]}(t) := \begin{bmatrix} v^{[i]}_{1}(t)& \omega_2^{[i]}(t)& \omega_3^{[i]}(t)\end{bmatrix}'. $
	Let $c^{[i]}(t)\in\mathbb{R}^3$ be a constant point in the body frame placed at a nonzero distance from the center of rotation of $\{B\}^{[i]}$ such that $c^{[i]}(t):=p^{[i]}(t) + R^{[i]}(t)\epsilon^{[i]}$, with  $\epsilon^{[i]}\in\mathbb{R}^3$. 
	
	Each vehicle is associated with a differentiable desired path $c_{d}^{[i]}:\mathbb{R}\to\mathbb{R}^3$ parametrized by $\gamma^{[i]}(t)\in\mathbb{R}$ and it is assumed that it can communicate (transmit and receive) a coordination parameter with the other vehicles according to a given graph $\mathcal{G}=(\mathcal{V},\mathcal{E})$ that defines the underlying communication topology. 
	
	The cooperative path-following (CPF) problem consists in designing motion control laws for the vehicles such that, as $t$ goes to infinity, the following holds:
	\begin{itemize}
		\item Convergence to the paths: The vectors $c^{[i]}(t)$ of the vehicles converge to $c_{d}^{[i]}(\gamma^{[i]}(t))$;
		\item Coordination: The network disagreement function 
		\begin{align}
		\phi(t):=\sum_{(i,j)\in \mathcal{E}}(\gamma^{[i]}(t)-\gamma^{[j]}(t))^2, \label{eq:disagreement}
		\end{align} 
		converges to the origin;
		\item Desired velocity along the path: The rate of the path parameters $\dot{\gamma}^{[i]}(t)$ converge to a desired constant value $v_d\in\mathbb{R}$.
	\end{itemize}
	
	One important difference with respect to the other works in the literature is that here we are looking for an optimization-based framework that integrates the path-following problem and the consensus problem with a combined cost function. Additional, we also consider input and output restrictions, where the output is the error that we would like to regulate to zero. More precisely, we define the output
    \begin{align}
	y^{[i]}(t)&=R^{[i]}(t)'(c^{[i]}(t)-c_{d}^{[i]}(\gamma^{[i]}(t))) \label{eq:yex}
	\end{align} 
	with $t\geq t_0$. 
	Notice that $y^{[i]}(t)=0$ if and only if $c^{[i]}(t)~=~c_{d}^{[i]}(\gamma^{[i]}(t))$. The rotation matrix is introduced to obtain an output with time derivative that directly depends on the input, as is made explicit in Section \ref{sss:aux}.
	
	For the numerical results described in Section \ref{sec:nres}, we present an illustrative scenario where three vehicles 
	are tasked to perform CPF along the geometric paths denoted in black in Fig. \ref{fig:pos}. In the desired configuration, the vehicles should move parallel to each other. Therefore, the path of each vehicle is parametrized such that if $\gamma^{[1]}=\gamma^{[2]}=\gamma^{[3]}$ then the desired positions $c_d^{[1]}(\gamma^{[1]})$, $c_d^{[2]}(\gamma^{[2]})$, and $c_d^{[3]}(\gamma^{[3]})$ are aligned as specified. The desired velocity of the parameter is chosen as $v_d = 2$. 
	The communication topology is set to be such that Vehicle 2, which is assigned to the path in the middle, communicates bidirectionally with Vehicles 1 and 3, but Vehicles 1 and 3 cannot communicate with each other.

	\vspace{-0.2cm}
	\subsection {Problem definition}
	The motivated CPF problem can be generalized as follows: Consider a set of $n_{\mathcal{I}}\in\mathbb{Z}_{\geq 1}$ agents (that is, dynamical systems, e.g., robotic vehicles) where the generic $i$-th agent, $i\in{\mathcal{I}}:= \{1,2,\ldots, n_{\mathcal{I}}\}$ is described by the continuous-time state equation
	\begin{subequations} \label{eq:allsys}
		\vspace{-1mm}\begin{align}
		\label{eq:ct_dynamic}
		\dot{x}^{[i]}(t)&=f^{[i]}(t,x^{[i]}(t),u^{[i]}(t)),\  x^{[i]}(t_0) = x_{0}^{[i]}, \  t\geq t_0
		\end{align}
		with $x^{[i]}(t)~\in~\mathbb{R}^{n^{[i]}}$ and $u^{[i]}(t)~\in~\mathcal{U}^{[i]}(t)\subseteq\mathbb{R}^{m^{[i]}}$ denoting the state vector and the input vector at time $t\geq t_0$, where $\mathcal{U}^{[i]}:\mathbb{R}_{\geq t_0}\rightrightarrows \mathbb{R}^{m^{[i]}}$ denotes the \emph{input constraint set}. The scalar $t_0\in\mathbb{R}$ and the vector $x_{0}^{[i]} \in~\mathbb{R}^{n^{[i]}}$ represent the initial time and state of the system, respectively, and the vector field $f^{[i]}~:~\mathbb{R}_{\geq t_0}~\times~\mathbb{R}^{n^{[i]}}~\times~\mathbb{R}^{m^{[i]}}~\rightarrow~\mathbb{R}^{n^{[i]}}$ can be different for each $i$.
		Consider also that to each agent there is an associated special variable $\gamma^{[i]}(t)\in \mathbb{R}^{n_c}$, named \emph{coordination vector}, that evolves with time according to the following dynamical model\begin{align}
		\label{eq:gammaDot}
		\dot{\gamma}^{[i]}(t)&=g(t)+u^{[i]}_{\gamma}(t),&& \gamma^{[i]}(t_0) = \gamma_{0}^{[i]}, && t\geq t_0
		\end{align}
		where ${g}:\mathbb{R}_{\geq t_0}\to \mathbb{R}^{n_c}$ is a generic function of time that is common to all the agents  and $u^{[i]}_{\gamma}(t)\in\mathbb{R}^{n_c}$ is an input signal. The agents communicate among each other according to a communication graph $\mathcal{G}:=(\mathcal{V},\mathcal{E})$, where the vertex set $\mathcal{V}$ collects all the indexes of the agents, that is, $\mathcal{V}={\mathcal{I}}$,  and the edge set $\mathcal{E}\subseteq \mathcal{V}\times \mathcal{V}$ is such that $(i,j)\in \mathcal{E}$ if and only if agent $i$ can access $\gamma^{[j]}(t)$. Therefore, agent $i$ can access from their neighborhoods $\mathcal{N}^{[i]}:=\{j:(i,j)\in \mathcal{E}\}$ at time $t$ the coordination vectors  $\gamma_{\mathcal{N}^{[i]}}(t):=\{\gamma^{[j]}(t):j\in \mathcal{N}^{[i]}\}$.
		
		The output of each system is defined as
		\begin{align}
		\label{eq:y}
		y^{[i]}(t)&=h^{[i]}(t,x^{[i]}(t),\gamma^{[i]}(t))\in\mathcal{Y}^{[i]}(t)
		\end{align}
		that is constrained within the \emph{output constrain set} denoted by $\mathcal{Y}^{[i]}~:~\mathbb{R}_{\geq t_0}~\rightrightarrows~\mathbb{R}^{p^{[i]}}$.
		
		Given the described setup, we can now formulate the following problem:
		\begin{prob}[Coordinated output regulation]\label{pm:coodrinated-output} 
			Design a control law for the input signals $\bu^{[i]}\in\mathcal{PC}(t_0,\infty)$ and $\bu^{[i]}_{\gamma}\in\mathcal{PC}(t_0,\infty)$, $i\in\mathcal{I}$, using only the information $x^{[i]}(t_k)$ and $\gamma_{\mathcal{N}^{[i]}}(t_k)$ available at the time instants $t_k\in\mathcal{T}:=~\{t_0, t_1,\dots \}$, such that for every $i\in {\mathcal{I}}$ the state vectors $x^{[i]}(t)$ and $y^{[i]}(t)$ are bounded for all $t\ge t_0$, and as time approaches infinity the following holds:
			\begin{enumerate}
				\item The output vector $y^{[i]}(t)\in \mathbb{R}^{p^{[i]}}$ converges to the origin;\label{item:i}
				\item \label{item:disag} The network disagreement function 
				defined in \eqref{eq:disagreement} 
				converges to the origin;\label{item:ii}
				\item The vector $\dot\gamma^{[i]}(t)$ converges to the predefined value ${g}(t)\in\mathbb{R}^{n_c}$, that is, $u^{[i]}_{\gamma}(t)$ converges to zero.\label{item:iii}\hfill $\square$
			\end{enumerate}
		\end{prob}
	\end{subequations}
	
	The problem addressed captures applications where it is important to solve in a unified way the output regulation problem and the consensus problem. In Problem \ref{pm:coodrinated-output}, the output is regulated to the origin without loss of generality. In fact, if a different output trajectory is desired, it is always possible to make a coordinate transformation with a new virtual output given by the difference between the original output and the desired one. Notice that for the particular case of ${g}(t)=0$, the design of a distributed control law for $u_\gamma^{[i]}$, with $i\in\mathcal{I}$
	corresponds to solving the standard consensus problem for single-integrator systems. Adding a term ${g}(t)\neq0$ in \eqref{eq:gammaDot}, common to all the vehicles, does not affect the value of the disagreement function, which only regards to the relative values, and allows the coordination parameters to converge to a (possibly time-varying) desired signal that is important in many applications, see, e.g., the motivating example, where $g(t)$ corresponds to the desired formation velocity $v_d$.
	\vspace{-0.2cm}
	\section{MPC Design and Convergence Analysis} \label{sec:main}
	This section proposes a sampled-data MPC approach with convergence guarantees that solves Problem~\ref{pm:coodrinated-output}. The key idea behind the proposed scheme is to define a performance index composed by a term for the output tracking MPC and another term that rewards the consensus among the coordination vectors. 
	
\begin{figure}[ttpb]
		\begin{center}
			\includegraphics[width=8.5cm]{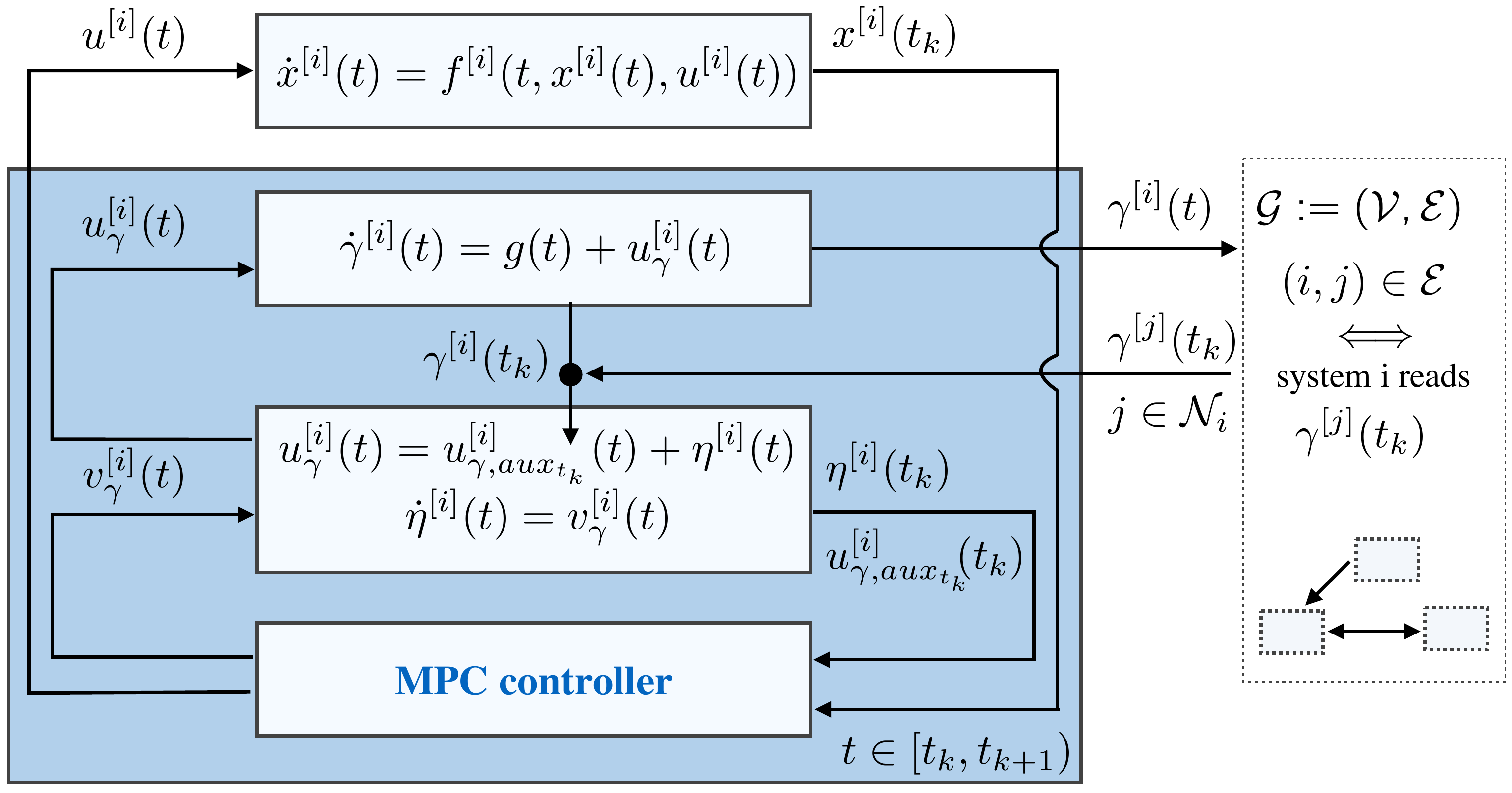}
			\vspace{-0.3cm}
			\caption{Block diagram of the proposed control architecture.}
			\label{fig:block}
		\end{center}
	\end{figure}

	
Fig. \ref{fig:block} shows a block diagram of the proposed control architecture, where the blue block corresponds to the algorithm that runs in each agent, which includes two internal states. The first one is the coordination vector $\gamma^{[i]}$ that evolves in time according to \eqref{eq:gammaDot} with $u^{[i]}_\gamma$ given by a continuous-time auxiliary consensus input signal $u^{[i]}_{\gamma,aux_t}$, derived in Section \ref{ss:auxcons}, plus an extra signal $\eta^{[i]}$, that is,
$
	u^{[i]}_{\gamma}(\tau)=u^{[i]}_{\gamma,aux_t}(\tau) + \eta^{[i]}(\tau).\nonumber
	$
	The second internal state $\eta^{[i]}\in\mathbb{R}^{n_c}$ evolves according to 
	\vspace{-1mm}\begin{align} 
	\dot{\eta}^{[i]}(t)=v^{[i]}_{\gamma }(t), && \eta^{[i]}(t_0)=\eta^{[i]}_0\label{eq:etaDot}
	\end{align}
	with initial condition $\eta_0^{[i]}\in\mathbb{R}^{n_c}$ and can be viewed as an error signal that denotes the difference between the auxiliary consensus control law computed in $u^{[i]}_{\gamma,aux_t}$ and the actual signal $u_\gamma^{[i]}$. This error signal is actuated through 
	$v^{[i]}_{\gamma}$ that is one of the outputs of the MPC controller. The other output is $u^{[i]}$ that directly commands the evolution  of the state vector $x^{[i]}$ according to \eqref{eq:ct_dynamic}.
	
	The MPC controller is a sampled-data scheme that solves at every time sample $t_k\in\mathcal{T}$ an optimization problem. 
	The obtained optimal inputs are applied open-loop to the system within the generic interval $t\in[t_k,t_{k+1})$, with $k\in\mathbb{Z}_{\geq 0}$, i.e., 
	\vspace{-1mm}\begin{align} 
	u^{[i]}(t)=\bar{u}^{\star\lfloor t \rfloor}(t),&&  v^{[i]}_\gamma(t)=\bar{v}^{\star\lfloor t \rfloor}_{\gamma }(t).\label{eq:kmpc}
	\end{align}
	The optimization is defined as follows.
	\begin{subequations} 
		\begin{defn}[Open-loop MPC problem]\label{eq:P} Given the tuple of parameters $p=(t,x^{[i]}, \gamma^{[i]},\gamma_{\mathcal{N}^{[i]}},\eta^{[i]}) \in \mathbb{R}_{\geq t_0} \times \mathbb{R}^{n^{[i]}} \times \mathbb{R}^{n_c} \times \mathbb{R}^{|\mathcal{N}^{[i]}|n_c}\times\mathbb{R}^{n_c}$, an horizon length $T\in\mathbb{R}_{>0}$, and the auxiliary signal $\bbaru^{[i]}_{\gamma,aux_t}\in\mathcal{C}(t,+\infty)$ that is introduced in \eqref{eq:ubaric}, the open-loop MPC optimization problem $\mathcal{P}(p)$ consists in finding the optimal control signals $\bbaru^{\star[i]} \in \mathcal{PC}(t,t+T)$ and $\bbarv_\gamma^{\star[i]}\in \mathcal{PC}(t,t+T)$ that solve
			\vspace{-1mm}\begin{align} 
			&J_T^{\star[i]}(p)= \min_{\substack{\bbaru^{[i]} \in \mathcal{PC}(t,t+T)\\ \bbarv_\gamma^{[i]}\in\mathcal{PC}(t,t+T)}}   J_T(p,\bbaru^{[i]},\bbarv_\gamma^{[i]}) \nonumber\\
			\text{s.t.} \quad & \dot{\bar{x}}^{[i]}(\tau)=f^{[i]}(\tau,\bar{x}^{[i]}(\tau),\bar{u}^{[i]}(\tau)),\quad\bar{x}^{[i]}(t) = x^{[i]},\nonumber\\
			\quad& \dot{\bar{\gamma}}^{[i]}(\tau)={g}(\tau)+\bar{u}^{[i]}_{\gamma}(\tau),\quad \bar{\gamma}^{[i]}(t)=\gamma^{[i]},\label{eq:dotgamma}\\
			\quad& \dot{\bar{\eta}}^{[i]}(\tau)=\bar{v}_\gamma^{[i]}(\tau),\quad \bar{\eta}^{[i]}(t)=\eta^{[i]},\label{eq:eta}\\
			\quad& \bar{y}^{[i]}(\tau)=h^{[i]}(\tau,\bar{x}^{[i]}(\tau),\bar{\gamma}^{[i]}(\tau)),\nonumber\\
			\quad& \bar{u}^{[i]}_{\gamma}(\tau)=\bar{u}^{[i]}_{\gamma,aux_t}(\tau) + \bar{\eta}^{[i]}(\tau),\label{eq:ubari}\\
			\quad& (\bar{u}^{[i]}(\tau),\bar{v}_\gamma^{[i]}(\tau),\bar{y}^{[i]}(\tau))\in \mathcal{U}^{[i]}(\tau)\times \mathcal{V}_\gamma^{[i]}(\tau)\times \mathcal{Y}^{[i]}(\tau)\nonumber\\
			\quad& (\bar{y}^{[i]}(t+T),\bar{\eta}^{[i]}(t+T))\in \mathcal{Y}_{aux}^{[i]}(t+T)\times \mathcal{B}(r_\eta^{[i]}) \nonumber\\
			\quad& |\bar{\eta}^{[i]}(\tau)|\leq  a_\eta^{[i]} e^{-\lambda_\eta^{[i]} (t-t_0)} \label{eq:ncon}
			\end{align} 
			where $\tau \in [t,t+T]$ and
			\vspace{-1mm}\begin{align}
			J^{[i]}_T&(p, \bbaru^{[i]},\bbarv_\gamma^{[i]}) 
			:=\int_t^{t+T} l^{[i]}(\tau,\bar{x}^{[i]},\bar{u}^{[i]},\bar{\gamma}^{[i]},\bar{u}^{[i]}_{\gamma})  d\tau \nonumber\\
			&+ m^{[i]}(t+T,\bar{x}^{[i]}(t+T),\bar{\gamma}^{[i]}(t+T))\nonumber\\
			&+ \int_t^{t+T}l_c^{[i]}(\bar{\eta}^{[i]},\bar{v}_\gamma^{[i]}) d\tau  +\frac{1}{2}m_\eta^{[i]}\big(\bar{\eta}^{[i]}(t+T)\big)^2 \label{eq:defJ}
			\end{align} \\[-3mm]
			and $\lambda_\eta^{[i]}>0$, $a_\eta^{[i]} \geq 0$, $m_\eta^{[i]} \geq 0$, $r_\eta^{[i]}\geq0$. 
			\hfill $\square$ 
		\end{defn}
	\end{subequations}
	
	From \eqref{eq:defJ}, one can see that the \emph{finite horizon cost} $J^{[i]}_T(\cdot)$, which corresponds to the \emph{performance index} of the MPC controller, is composed of two stage costs, i.e., the \emph{regulation stage cost} $l^{[i]}~:~\mathbb{R}_{\geq t_0}~\times~\mathbb{R}^{n^{[i]}}~\times~\mathbb{R}^{m^{[i]}}\times \mathbb{R}^{n_c}\times \mathbb{R}^{n_c}~\rightarrow~\mathbb{R}_{\geq 0} $ and the \emph{consensus stage cost} $l^{[i]}_c~:~\mathbb{R}^{n_c}\times \mathbb{R}^{n_c}  \rightarrow~\mathbb{R}_{\geq 0}$, and two terminal costs, i.e., the \emph{regulation terminal cost} $m^{[i]}~:~\mathbb{R}_{\geq t_0} \times \mathbb{R}^{n_c} \times \mathbb{R}^{n_c}~\rightarrow~\mathbb{R}_{\geq 0}$, and the \emph{consensus terminal cost} ${1\over 2}m_\eta^{[i]}\big(\bar{\eta}^{[i]}(t+T)\big)^2$. 
	The key idea to combine these two stage costs is to optimally balance the output regulation objective ($\bar{y}^{[i]}(\tau)=0$) and the consensus objective ($\bar{\eta}^{[i]}(\tau)=0$) according to the weights specified in the performance index. To ensure that this difference given by $\eta$ converges to zero, we include the constraint \eqref{eq:ncon}, which also implies a specific exponentially  bound. Although, in practice, taking high values of $a_\eta^{[i]}>0$ and small values of $\lambda_\eta^{[i]}>0$, such constraint is never active and the convergence rate of $\bar{\eta}^{[i]}$ is fully determined by the performance index. Furthermore, constraint \eqref{eq:ncon} can be omitted if the vehicle is disconnected from the network ($\mathcal{N}^{[i]}=\emptyset$ and $i\notin\mathcal{N}^{[j]}$ with $j=\mathcal{I}$, $i\neq j$), in which case the scheme results in a pure output regulation controller.
	The regulation terminal cost is defined over the set of $(t,x^{[i]},\gamma^{[i]})$ such that the associated $y^{[i]}(t)\in\mathcal{Y}_{aux}^{[i]}(t)$ belongs to the \emph{regulation terminal set} $\mathcal{Y}_{aux}^{[i]}:\mathbb{R}_{\geq t_0}\rightrightarrows \mathbb{R}^{p^{[i]}}$. Similarly, the consensus terminal cost is evaluated with $\bar{\eta}^{[i]}(t+T)$ constrained in the \emph{consensus terminal set} $\mathcal{B}(r_\eta^{[i]})$. 
	
	The main result of this section follows next. For the sake of clarity, the assumptions and the associated design methods are reported in Sections \ref{ss:desass} and \ref{ss:auxcons}.
	\begin{thm} \label{th:main}
		Consider the set of constrained dynamical systems \eqref{eq:allsys} that communicates according to a communication network $\mathcal{G}=(\mathcal{V},\mathcal{E})$ as described in Section~\ref{sec:problem_definition}. 
		If Assumptions~\ref{ass:1}-\ref{ass:con} hold, then the proposed sampled-data MPC control law \eqref{eq:kmpc} solves Problem~\ref{pm:coodrinated-output}. The region of attraction of the proposed controller corresponds to the set of initial conditions of the system such that the open-loop MPC problem is feasible. 
	\end{thm}
	\begin{rem}
			The knowledge of the state of the neighboring systems of agent $i$ is indirectly captured by their coordination vectors. For instance, in the cooperative path-following example, the coordination vector $\gamma^{[i]}$ defines where the desired position is within the desired path, i.e., $c_d^{[i]}(\gamma^{[i]})$, and formation is achieved when all the coordination vectors are in consensus (and the associated path-following errors are zero). Because of this, the knowledge of the dynamical model of the neighboring agents, their state, or prediction of the state, is not required, and only the value of their coordination vectors is shared. In addition, it is worth noticing that the controller only shares the \emph{current} value of the coordination vector, and not its prediction, because the proposed scheme builds on a pre-existing auxiliary consensus control law that only requires the current values. The MPC controller modifies such auxiliary control law with an additive vanishing term that is used to optimally balance the coordination objective with the path-following objective.  \end{rem}
\vspace{-0.3cm}
	\subsection{Main assumptions}\label{ss:desass}

	Whenever it is clear from the context, we omit the explicit superscript $[i]$ to improve the clarity of the paper. The first set of conditions is obtained by adapting the standard MPC sufficient conditions for state convergence to the origin in the state space, to the convergence of the output to the origin in the output space (see, e.g., \cite{Alessandretti2016} with Remark 18 for the generic case of time-varying system considered in this paper). 
	
	\begin{assum}\label{ass:1} The input constraint set $\;\mathcal{U}(t)$ is compact for all $t\geq t_0$ and uniformly bounded over time. Moreover, the function $f(\cdot)$ in \eqref{eq:ct_dynamic} is locally Lipschitz in $x$, piecewise continuous in $t$ and $u$, and bounded for bounded $x$ in the region of interest, i.e., the set $\{ \|f(t,x,u)\|: t\geq t_0, x\in \bar{\mathcal{X}}, u\in \mathcal{U}(t)\}$ is bounded for any bounded $\bar{\mathcal{X}}\subset \mathbb{R}^n$.
		\hfill $\square$\end{assum}
	
	\begin{assum}\label{ass:1b} The function $g:\mathbb{R}_{\geq t_0}\to\mathbb{R}^{n_c}$ in \eqref{eq:gammaDot} is uniformly bounded over time.\hfill $\square$\end{assum}
	
	To guarantee that the state trajectories will be uniformly bounded over time provided that the output and input trajectories are bounded, the following assumption is considered.
	
	\begin{assum} \label{ass:y} Consider the output defined in  \eqref{eq:y}. 
		\begin{enumerate}
			\item  The system \eqref{eq:allsys} satisfies for all $t\geq t_0$ the condition
			\vspace{-1mm}\begin{align}
			\|x(t)\|&\leq\beta_x(\|x_0\|,t-t_0)+\sigma_u(\|u\|_{[t_0,t)})\nonumber\\
			&\quad 
			+\sigma_y(\|y\|_{[t_0,t)}) +\sigma_x \label{eq:issy}
			\end{align} 
			for a class-$\mathcal{KL}$ function $\beta_x~:\mathbb{R}_{\geq0}~\times~\mathbb{R}_{\geq0}~\to~ \mathbb{R}_{\geq0}$, class-$\mathcal{K}$ functions $\sigma_u,\sigma_y~:~\mathbb{R}_{\geq0}~\to~\mathbb{R}_{\geq0}$, and a constant scalar $\sigma_x\geq0$ that could be dependent on the initial condition $x_0$;
			\item The gradient of the right-hand side of the output equation \eqref{eq:y} $\nabla h(t,x,\gamma)$ is uniformly bounded over time for bounded values of $x$, i.e., for all $x$ with $\|x\|\leq B$, there exists a scalar $b_B>0$, possibly dependent on $B\geq0$
			such that 
			$
			\|\nabla h(t,x,\gamma)\|\leq b_B 
			$ for all $t\geq t_0$ and $\gamma\in\mathbb{R}^{n_c}$. \label{item:yDotB}\hfill $\square$
		\end{enumerate}
	\end{assum}
	
	Condition \eqref{eq:issy} is used to avoid cases where the system has, e.g., non-observable unstable states. More specifically, for the case $\sigma_x=0$,  inequality \eqref{eq:issy} corresponds to the input-output-to-state stability (IOSS) condition (see, e.g., \cite{Sontag1999,Krichman2001}) that represents a detectability condition of the state from the output (see also \cite{Prieur2011} and Proposition 2.6 of \cite{Krichman2001} for the case of linear systems). It is worth noticing that this condition does not imply a-priori that system \eqref{eq:ct_dynamic} needs to be stable (it can be indeed open loop unstable). It is however possible to conclude stability of the closed-loop (or more precisely internal stability) provided that one can show that the closed-loop system is external stable (see \cite{Sontag06inputto} where it is shown the equivalence between the IOSS and the input-to-output stability (IOS) with the input-to-state stability (ISS) property). Thus, only after properly regulating the output it is possible to guarantee the boundedness of the state trajectory. In this paper, we further relax this assumption allowing $\sigma_x\ge 0$. Broadly speaking, \eqref{eq:issy} requires the state of the system to be uniformly bounded for uniformly bounded output and input signals. Similarly, also the item~\ref{item:yDotB} of the latter assumption is rather general, and we refer to the illustrative example for more insight on when such conditions hold.
	
	In a classic MPC scheme, to guarantee closed-loop convergence of the state trajectory to the origin, the performance index and the constraint sets of the open-loop MPC problem are required to satisfy a set of sufficient conditions (see, e.g., \cite{CHEN1998,Fontes2001,Findeisen2003}).  In the following two assumptions, similar sets of conditions are required to guarantee that $y$ and $\eta$, respectively, converge to the origin.
	
	\begin{assum}\label{sc} 
		The following hold:
		\begin{enumerate}
			\item\label{item:sc1} The output constraint set $\mathcal{Y}(t)$ and the terminal set $0~\in~\mathcal{Y}_{aux}(t)~\subseteq~\mathcal{Y}(t)$ are closed, connected, and contain the origin for all $t\geq t_0$;
			\item\label{ascii} There is a class-$\mathcal{K}_\infty$ function $\alpha_s:\mathbb{R}_{\geq 0}~\rightarrow~\mathbb{R}_{\geq 0}$ such that $
			l(\tau,\bar{x},\bar{u},\bar{\gamma},\bar{u}_{\gamma})~\geq~\alpha_s(\|y\|)$
			for all $(\tau,\bar{x},\bar{u},\bar{\gamma},\bar{u}_{\gamma}) \in \mathbb{R}_{\geq t_0} \times \mathbb{R}^{n} \times \mathcal{U}(\tau)\times \mathbb{R}^{n_c}\times \mathbb{R}^{n_c}$;
			\item For any given values of $(x,u,\gamma,u_{\gamma}) \in \mathbb{R}^{n} \times \mathbb{R}^{m}\times \mathbb{R}^{n_c}\times \mathbb{R}^{n_c}$ the functions $l(t,x,u,\gamma,u_{\gamma})$ and $m(t,x,\gamma)$ are uniformly bounded over time $t\geq t_0$;\label{item:sc1auxs}
			\item \label{item:aux}There exists a auxiliary regulation control law 
			$k_{aux}~:~\mathbb{R}_{\geq t_0}~\times~\mathbb{R}^{n}~\times ~\mathbb{R}^{n_c}~\times~\mathbb{R}^{n_c}~\rightarrow~\mathbb{R}^{m}$ such that, for the associated closed-loop system \eqref{eq:ct_dynamic} with $u(t)=k_{aux}(t,x,\gamma,u_\gamma)$, \eqref{eq:gammaDot} with $u_\gamma(t) = \eta(t)$, and \eqref{eq:etaDot} with $v_\gamma(t)=-\lambda_{\eta}\eta(t)$, $\lambda_{\eta}>0$, with initial time and states $(\hat{t},\hat{x},\hat{\gamma},\hat{\eta})\in \mathbb{R}_{\geq t_0+T}\times \mathbb{R}^n\times \mathbb{R}^{n_c}\times \mathbb{R}^{n_c}$, for all $t\geq t_0+T$ the input and output vectors satisfy $u(t)~\in~\mathcal{U}(t)$ and $y(t)~\in~\mathcal{Y}_{aux}(t)$ and the condition 
			\vspace{-1mm}\begin{align}
			&m(\hat{t}\!+\!\delta,x(\hat{t}\!+\!\delta),\gamma(\hat{t}\!+\!\delta))\!-\!m(\hat{t},x(\hat{t}),\gamma(\hat{t})) \!\leq \!\!- \!\int_{\hat{t}}^{\hat{t}+\delta} \!\!\!\!\!\!\!\!\!l(t) dt\label{eq:decrese}
			\end{align}\\[-4mm]
			holds for any $\delta > 0$.\label{item:sc1aux}\hfill $\square$
	\end{enumerate}\end{assum}
	
	Assumption \ref{sc}.\ref{item:aux} requires the existence of an auxiliary control law able to steer the system to track the output trajectory obtained when $\gamma(t)$ is driven by the (dynamic) controller $u_\gamma(t)=\eta(t)$,  
	$\dot\eta(t) = v_\gamma(t)=-\lambda_{\eta}\eta(t)$. The design of a suitable terminal set and terminal cost follows analogously to classic MPC, where the auxiliary controller drives the system to the origin, instead of driving the output to a trajectory. Specifically, if $k_{aux}(\cdot)$ is given with a global Lyapunov function $V_{aux}:\mathbb{R}_{\geq t_0+T}\times\mathbb{R}^{p}\times\mathbb{R}^{n_c}\to\mathbb{R}_{\geq 0}$ such that $\alpha_{aux,1}(\|y\|)\leq V_{aux}(t,x,\gamma) \leq \alpha_{aux,2}(\|y\|)$, $\alpha_{aux,3}(\|y\|) \geq l(\tau,x,k_{aux}(\cdot),\gamma,u_{\gamma})$, and $\dot{V}_{aux}(t,x,\gamma) \leq - \alpha_{aux,3}(\|y\|)$
	for three class-$\mathcal{K}_{\infty}$ functions $\alpha_{aux,1}(\cdot)$, $\alpha_{aux,2}(\cdot)$, and $\alpha_{aux,3}(\cdot)$, then $m(t,x,\gamma)=V_{aux}(t,x,\gamma)$ and $\mathcal{Y}_{aux}=\mathbb{R}^{p}$ is a valid choice of terminal cost and terminal set. Moreover, if the conditions on $V_{aux}$ only hold in a neighborhood $\mathcal{N}$ around $y=0\in \interior(\mathcal{N})$, Assumption \ref{sc}.\ref{item:aux} is satisfied by choosing the set $\mathcal{Y}_{aux}(\cdot)$ to be a level set of $m(\cdot)$ (therefore positively-invariant) contained in $\mathcal{N}$. Notice that, although for the case of differentiable terminal costs \eqref{eq:decrese} can be rewritten as $\dot{m}(t,x,\gamma)\leq - l(t,\bar{x},k_{aux}(\cdot),\bar{\gamma},\bar{u}_{\gamma})$, we adopt the integral representation to allow non-differentiable terminal cost, as it is the case of the illustrative example in this paper.
	
	\begin{assum}\label{sc2}
		The following hold:
		\begin{enumerate}
			\item The constraint set $\mathcal{V}_\gamma(t)$ is compact, uniformly bounded over time and such that $\mathcal{B}(r_\eta\lambda_\eta)\subseteq\mathcal{V}_\gamma (t)$, for all $t\geq t_0$;\label{item:sc11}
			\item\label{asciii} The consensus cost function $l_c(\cdot)$ is zero with $\eta~=~0$ and the function $\alpha_c:\mathbb{R}_{\geq 0} \rightarrow \mathbb{R}_{\geq 0}$ is such that $l_c(\eta,v_\gamma)~\geq\alpha_c(\|\eta\|)$ for all $(\eta,v_\gamma)~\in~\mathbb{R}^{n_c}\times\mathbb{R}^{n_c}$, and $m_\eta\lambda_\eta \eta^2\geq l_c(\eta,-\lambda_{\eta}\eta)$. \hfill $\square$
		\end{enumerate}
		\end{assum}
	
	Similarly to \eqref{item:sc1aux} in Assumption \ref{sc}, Assumption \ref{sc2} implies that, for the closed-loop \eqref{eq:etaDot} with $v_\gamma = -\lambda_{\eta}\eta$, the consensus terminal cost ${1\over 2}m_\eta\eta^2$ satisfies 
	\vspace{-1mm}\begin{align}
	{d\over dt} {1\over 2}m_\eta\eta^2 = - m_\eta\lambda_{\eta}\eta^2 \leq - l_c(\eta,-\lambda_{\eta}\eta)\label{eq:inmcon}
	\end{align}
	that is used in the convergence analysis of the proposed controller.
	
	The proposed MPC controller balances the regulation objective (Problem \ref{pm:coodrinated-output}, item \ref{item:i}) with the consensus objective (Problem \ref{pm:coodrinated-output} items \ref{item:disag} and \ref{item:iii}). Since such objectives are generally conflicting in a transient phase, the performance index of the MPC controller can be tuned to favorite either output regulation or consensus.  The existence of the auxiliary control law in Assumption \ref{sc}.\ref{item:aux} implicitly implies that, asymptotically, both objectives can be satisfied (i.e., $\phi(t)=0$ and $y^{[i]}(t)=0$, $i\in\mathcal{I}$, is a feasible solution for system \eqref{eq:allsys} for all $t \geq t_0$).

	\begin{assum}\label{sc3} Consider the open-loop MPC optimization problem from Definition~\ref{eq:P}. The regulation stage cost $l(\cdot)$ and the regulation terminal cost $m(\cdot)$ are Lipschitz continuous on $(\gamma,u_\gamma)$, i.e., there exists a pair of constants $C_l\geq 0$ and $C_m\geq 0$ such that 
		\vspace{-1mm}\begin{align} 
		&|l(t,x,u,\gamma_1,u_{\gamma,1})-l(t,x,u,\gamma_2,u_{\gamma,2})| 
		\leq C_l \left\|\begin{bmatrix}\gamma_1-\gamma_2\\ u_{\gamma,1}-u_{\gamma,2}\end{bmatrix}\right\|\nonumber\\
		&|m(t,x,\gamma_1)-m(t,x,\gamma_2)|
		\leq C_m \left\|\begin{bmatrix}\gamma_1-\gamma_2 \\ u_{\gamma,1}-u_{\gamma,2}\end{bmatrix}\right\|\nonumber
		\end{align}
		holds for any given $(t,x,u)\in \mathbb{R}_{\geq t_0}\times \mathbb{R}^{n}\times \mathbb{R}^{m}$, $(\gamma_1,u_{\gamma,1})\in\mathbb{R}^{n_c}\times \mathbb{R}^{n_c}$, and $(\gamma_2,u_{\gamma,2})\in\mathbb{R}^{n_c}\times \mathbb{R}^{n_c}$. \hfill $\square$
	\end{assum}
	
	In Assumption~\ref{sc3}, the functions  $m(\cdot)$ and $l(\cdot)$ are required to be Lipschitz only on the variables $(\gamma,u_\gamma)$, which makes the assumption rather general. Moreover, since the Lipschitz constants are not used in the design phase, only their existence is required, and not their computation.
	\vspace{-0.3cm}
	\subsection{Continuous-time auxiliary consensus input signal}\label{ss:auxcons}
	
	To design the continuous-time auxiliary consensus input signal, we rely on a class of consensus control laws for discrete-time systems. In particular, we assume the following:
	
	\begin{assum}[Auxiliary consensus control law]\label{ass:con} Consider a set of discrete-time systems that communicate according to the same communication graph $\mathcal{G}~=~(\mathcal{V},\mathcal{E})$ introduced in Section~\ref{sec:problem_definition}, and satisfies for each $i\in{\mathcal{I}}$ the dynamical model
		\vspace{-1mm}\begin{align}
		\xi^{[i]}(k+1)=\xi^{[i]}(k)+ k_{con}(\xi^{[i]}(k),\xi_{\mathcal{N}^{[i]}}(k))+\eta^{[i]}(k)\label{eq:conSys}
		\end{align}
		where $\xi^{[i]}(k)\in \mathbb{R}^{n_c}$, $\eta^{[i]}(k)\in \mathbb{R}^{n_c}$, and $\xi_{\mathcal{N}^{[i]}}(k)$ denote the $i$-th coordination vector, an external vector, and the coordination vectors from the neighborhood $\mathcal{N}^{[i]}$, respectively, at step $k~\in~\mathbb{Z}_{\geq k_0}$, with $\xi^{[i]}(k_0) = \xi_{0}^{[i]}\in\mathbb{R}^{n_c}$ denoting the initial condition of the system $i$ at the initial time step $k_0\in\mathbb{Z}$. We assume that there exists an auxiliary consensus control law $k_{con}(\cdot)$ that satisfies the property that if
		\vspace{-1mm}\begin{align}
		\|\eta^{[i]}(k)\|\leq a^{[i]}_\eta e^{-\lambda^{[i]}_\eta (k-k_0)}\label{eq:boundeta}
		\end{align} 
		for some constants $\lambda_\eta>0$ and $a_\eta\geq0$ then:
		\begin{enumerate}
			\item \label{item:contozero}As $k\to \infty$, the discrete disagreement function $\phi(k)~=~\sum_{(i,j)\in \mathcal{E}}(\xi^{[i]}(k)-\xi^{[j]}(k))^2$  converges asymptotically to zero;
			\item \label{item:conbounded}The consensus control law $k_{con}:\mathbb{R}\times \mathbb{R}^{|\mathcal{N}^{[i]}|} \to \mathcal{U}_c$, is bounded, i.e.,  $\mathcal{U}_c\subset \mathbb{R}^{n_c}$ is bounded;
			\item \label{item:betaint} There exists an integrable class-$\mathcal{KL}$ function  $\beta:\mathbb{R}_{\geq0}\times \mathbb{R}_{\geq0}\to \mathbb{R}_{\geq0}$ such that
			\vspace{-1mm}\begin{align*}
			\|k_{con}(\xi^{[i]}(k),\xi_{\mathcal{N}^{[i]}}(k))\|\leq \beta(\sum_{i\in\mathcal{I}}\|\xi_0^{[i]}\|,k-k_0). 
			\end{align*}\vspace{-9mm}\\
		\end{enumerate}
	\end{assum}
	Assumption \ref{ass:con} requires a distributed consensus law that robustly solves (in the sense of the required properties mentioned above) the consensus problem for multi-agent systems, where each agent is given by a simple discrete integrator. Moreover, it implicitly defines the class of network topologies that can be addressed by the proposed scheme. Therefore, in the design phase, an auxiliary consensus control law should be chosen accordingly to the specific application/network topology to satisfy Assumption~\ref{ass:con}. See Appendix \ref{app:analycon} for a brief background on the analysis of consensus control laws.
	
	In particular, for the case of time-invariant balanced strongly connected graphs, we show that the consensus control law 
	\vspace{-1mm}\begin{align}
	k_{con}(\xi^{[i]},\xi_{\mathcal{N}^{[i]}})= -\bar{\epsilon} \sum_{j\in\mathcal{N}^{[i]}}a_{ij}(\xi^{[i]}-\xi^{[j]}) \label{eq:conscon},
	\end{align} \\[-0.4cm]
	for a suitable choice of the constants $a_{ij}>0$ and $\bar{\epsilon}$, satisfies Assumption~\ref{ass:con}.  
	\begin{prop}\label{lem:assat}
		Consider the network of discrete-time systems in Assumption~\ref{ass:con} with  $n_c=1$ (scalar state) and where $\mathcal{G}$ is a balanced strongly connected graph. Then, the consensus control law \eqref{eq:conscon} with $\bar{\epsilon} \in (0,1/\Delta)$ where $\Delta = \max_{i}(\sum_{j\neq i}a_{ij})$, satisfies Assumption~\ref{ass:con}.  \hfill $\square$
	\end{prop}
	
	Next, the discrete auxiliary consensus control law $k_{con}(\cdot)$ required by Assumption \ref{ass:con} is used to design a continuous-time sample-data consensus control signal for \eqref{eq:gammaDot} that only uses state observations taken at the discrete time instants
	$\mathcal{T}~=~\{t_0, t_1,\dots \}$
	where it is assumed that the difference between time samples $\delta_k:=t_{k+1}-t_k$ is upper bounded by the horizon length $T$ of the MPC controller, and lower bounded by a fixed value $\delta_{lb}> 0$.   
	
	\begin{prop}\label{pro:new}
		Consider the discrete-time auxiliary consensus control law in Assumption \ref{ass:con}, the continuous-time system \eqref{eq:gammaDot}, and the continuous-time auxiliary consensus input signal $\bbaru^{[i]}_{\gamma,aux_{t}}\in\mathcal{C}(t,+\infty)$ defined as 
		\vspace{-1mm}\begin{align}
		\bar{u}^{[i]}_{\gamma,aux_{t}}(\tau) &:= 
		{\begin{cases}
		{1\over \delta_k}k_{con}(\gamma^{[i]}(t_k),\gamma_{\mathcal{N}^{[i]}}(t_k)),&  \!\!\tau \in \![t,t+\delta_k] \\
		0,& \!\!\tau\! > \!t\!+\!\delta_k
		\end{cases}}\nonumber\\[-5mm]\label{eq:ubaric}
		\end{align}
		with $t\in [t_k,t_{k+1})$ and $\delta_k = t_{k+1}-t_k$, $k\in\mathbb{N}_{\geq 0}$. Then, for the closed-loop system \eqref{eq:gammaDot} with $u^{[i]}_\gamma(t)=\bar{u}^{[i]}_{\gamma,aux_{\lfloor t \rfloor}}(t)$ , the disagreement function \eqref{eq:disagreement} converges asymptotically to zero.  \hfill $\square$
	\end{prop}
	
	\section{Cooperative path following for multiple 3D nonholonomic vehicles}\label{sec:numerical}
	
	In this section, we apply the proposed optimization-based control framework to solve the CPF problem presented in Section 
	\ref{sec: CPF robot}. Numerical simulations are then provided to illustrate the effectiveness of the proposed approach. 
	
	\vspace{-0.3cm}
	\subsection{Control Design}
	
	In this section, we show how to satisfy all the assumptions presented in Section \ref{sec:main} for the CPF problem for nonholonomic robotic vehicles of Section \ref{sec: CPF robot}. Indeed, the latter can be formulated in the form of Problem \ref{pm:coodrinated-output} by choosing $x^{[i]}(t) = \begin{pmatrix}p^{[i]}(t)&r(R^{[i]}(t))'\end{pmatrix}'$, with associated dynamical model from \eqref{eq:model}, where $r(R^{[i]}(t))$ denotes a vectorial parameterization of the matrix $R^{[i]}(t)$, e.g., the components of the rotation matrix, and with output $y^{[i]}(t)$ given in \eqref{eq:yex}. It is worth noticing that despite this example uses homogeneous systems, Theorem \ref{th:main} applies to systems with possibly different dynamical models. The reference paths $c_d^{[i]}$ are designed to be bounded since we would like to guarantee boundedness of the state trajectories of the vehicles. The control design consists of six steps: 
	
	\subsubsection{Auxiliary consensus control law}
	We propose  the consensus control law \eqref{eq:conscon} with $a_{ij}=1$ for $(i,j) \in \{(1,2), (2,1), (2,3), (3,2)\}$ and zero otherwise, and $\bar{\epsilon}=0.0125\!<\!{1\over \Delta_d}$. Thus, Assumption~\ref{ass:con} is satisfied by Proposition~\ref{lem:assat}. 
	
	\subsubsection{Auxiliary regulation control law}\label{sss:aux} We consider a saturated version of the auxiliary control law mentioned in \cite{Alessandretti2013}. From \eqref{eq:model} and defining $c_{d_\gamma}(\gamma)={\partial \over \partial \gamma}c_d(\gamma)$, the time derivative of the output \eqref{eq:yex} is
	\vspace{-1mm}\begin{align}
	\dot{y}&=-\Omega(\omega) R'(p-c_d)+R'(\dot{p}-\dot{c_d})\nonumber\\
	&=-\Omega(\omega) y +\Omega(\omega) \epsilon +\begin{bmatrix}v_1&0&0\end{bmatrix}'-R'\dot{c_d}\nonumber\\
	&= -\Omega(\omega) y -R'c_{d_\gamma}(\gamma)\dot{\gamma}+\Delta u,\quad \Delta:={\tiny \begin{bmatrix}1&\epsilon_3&-\epsilon_2\\
		0&0&\epsilon_1\\
		0&-\epsilon_1&0\end{bmatrix}}\nonumber
	\end{align} 
	where we used the fact that $\Omega(\omega) \epsilon =-\Omega(\epsilon)\omega$. Then, for any choice of $\epsilon$ with $\epsilon_1\neq0$ the input $u=k_{aux}(t,x,\gamma,u_\gamma)$ with
	\vspace{-1mm}\begin{align}
	k_{aux}(\cdot) = 
	\begin{cases}
	\Delta^{-1}\left( R'{\partial \over \partial \gamma}c_d(\gamma)(v_d+u_\gamma)-K {y \over \|y\|}	\right),& \|y\| \neq 0\\
	\Delta^{-1} R'{\partial \over \partial \gamma}c_d(\gamma)(v_d+u_\gamma),& \|y\| = 0
	\end{cases} \label{eq:exAux}
	\end{align}  \\[-3mm]
	results in 
	$\dot{y} = 
	-\Omega y-K {y \over \|y\|}$, for  $\|y\| \neq 0$, and 
	$\dot{y}=0$, for $\|y\| = 0$,
	where the matrix $K\succ 0$ is a tuning parameter. Therefore, the Lyapunov-like function $W = \|y\|$ satisfies
	\vspace{-1mm}\begin{align} 
	\dot{W}=
	\begin{cases}
	{y'\dot{y}\over \|y\|}=-{y'Ky\over \|y\|^2}\leq-\lambda_{min}(K),&  \|y\| \neq 0\\
	0,&  \|y\| = 0,
	\end{cases} \nonumber
	\end{align} 
	where we used the fact that $\Omega$ is skew-symmetric and therefore $y'\Omega y=0$ for all $y\in\mathbb{R}^3$. As a consequence, the vector $y$ converges in finite time to the origin as follows
	\vspace{-1mm}\begin{align}
	\|y(\tau)\|\leq 
	\begin{cases}
	\|y(t)\| - \lambda_{min}(K)(\tau-t),& \tau\in [t,\bar{t}] \\
	0, & \tau> \bar{t}
	\end{cases}\label{eq:bountOny}
	\end{align}
	with $\bar{t}:= t+{\|y(t)\|\over\lambda_{min}(K)}$. This controller is used in the following sections for the computation of the stage cost, terminal cost, and terminal constraint set. The values used in the numerical simulation are $\epsilon^{[i]}\!=\![-0.5, 0, 0]'$ for all $i\!\in\!\mathcal{I}$ and $K\!=\!0.2I_{3\times 3}$.
	%
	\subsubsection{Regulation stage cost}  Assumption~\ref{sc}.\ref{ascii} is satisfied by the regulation stage cost 
	\vspace{-1mm}\begin{align}
	l(t,x,u,\gamma,u_\gamma)~=~\|y\|^2_Q + \|u-k_{aux}(t,x,\gamma,u_\gamma)\|^2_U\label{eq:exl}
	\end{align} for any $Q\succ0$ and $U\succeq0$ of suitable dimensions. In the simulations, we use $U=I_{3\times3}$. To highlight the behavior of the proposed scheme, $Q$ is first selected as $Q=0.1I_{3\times3}$ and then updated to $Q=100I_{3\times3}$ as described in Section~\ref{sec:nres}.
	
	\subsubsection{Regulation terminal cost} Let $(\by_{aux},\bu_{aux})$ be the pair of output and input trajectories of the system in closed-loop with the auxiliary regulation control law \eqref{eq:exAux}, starting at the time and output pair $(\hat{t},\hat{y})$. Then, using \eqref{eq:bountOny} in  \eqref{eq:exl}, it follows that the associated regulation stage cost is upper bounded as 
	$l(\tau,y_{aux},u_{aux})\leq\hat{l}(\tau;\hat{t},\hat{y})$, with
	\vspace{-1mm}\begin{align}
	\hat{l}(\cdot):=
	\begin{cases}
	\lambda_{max}(Q)(\|\hat{y}\| - \lambda_{min}(K)(\tau-t))^2, \tau\leq  \hat{t}+\tfrac{\|\hat{y}\|}{\lambda_{min}(K)}\nonumber \\
	0, \qquad\qquad\qquad\qquad\qquad\qquad\qquad \tau> \hat{t}+{\|\hat{y}\|\over\lambda_{min}(K)}\nonumber 
	\end{cases}
	\end{align}
	where $\hat{l}(\cdot)$ satisfies
	$
	\hat{l}(\tau;\hat{t}+\delta,y_{aux}(\hat{t}+\delta))\leq \hat{l}(\tau;\hat{t},\hat{y})$ for all $\delta \geq 0$ and $
	\lim_{\tau\to\infty} \hat{l}(\tau;\hat{t},\hat{y})  = 0\nonumber
	.$ We can now conclude by Lemma 24 of \cite{Alessandretti2016} that the regulation terminal cost 
	$
	m(\hat{t},\hat{y})=\int_{\hat{t}}^{+\infty} \hat{l}(\tau;\hat{t},\hat{y})d\tau
	= \left. -\tfrac{\lambda_{max} (Q)}{ 3 \lambda_{min}(K)}{(\|\hat{y}\| - \lambda_{min}(K)\tau)^3}\right|_{\tau=0}^{\|\hat{y}\|\over\lambda_{min}(K)}
	=  \tfrac{\lambda_{max} (Q)}{ 3 \lambda_{min}(K)}\|\hat{y}\|^3\label{eq:exm} 
	$
	satisfies the terminal cost decrease of Assumption~\ref{sc}.\ref{item:sc1aux}. 
	
	\subsubsection{Regulation terminal set} 
	From \eqref{eq:exAux} we have
	\begin{subequations}\label{eq:bounduaux}
		\vspace{-1mm}\begin{align}
		\|[u_{aux}(\tau)]_1\| \leq \left\| [\Delta^{-1}]_1\right\| \bar{n} + \|[\Delta^{-1}K]_1\| &=:v_{max}\\
		\|[u_{aux}(\tau)]_2\| \leq \left\| [\Delta^{-1}]_2\right\| \bar{n} + \|[\Delta^{-1}K]_2\| &=:\omega_{2,max}
		\\
		\|[u_{aux}(\tau)]_3\| \leq \left\| [\Delta^{-1}]_3\right\| \bar{n} + \|[\Delta^{-1}K]_3\| &=:\omega_{3,max}
		\end{align}
	\end{subequations}
	where  $\bar{n}:=(\|v_d\| + r_\eta )\sup_{\gamma}\left\|c_{d_\gamma}(\gamma)\right\|$. Using now the fact that the derivative of the desired path is required to be bounded, then
	 from \eqref{eq:bounduaux}, by selecting an uniformly bounded $\mathcal{U}(t)$ with
	\begin{equation*}
	\left\{v,\omega_2,\omega_3 \!:\! |v_1|\leq \!v_{max}, |\omega_2|\leq\! \omega_{2,max}, |\omega_3|\leq \!\omega_{3,max}\right\}\!\subseteq \!\mathcal{U}(t)\nonumber
	\end{equation*}
	one conclude that the auxiliary regulation controller is always feasible, and therefore the regulation terminal set can be omitted, i.e., choosing $\mathcal{Y}_{aux}(t)=\mathbb{R}^3$ for all $t \geq t_0$. The selected regulation terminal set, regulation terminal cost, and constraint sets satisfy Assumption~\ref{sc}, items~\ref{item:sc1} and \ref{item:sc1aux}.
	
	The terminal constraint on $\bar{\eta}(t+T)$ is chosen with $r_{\eta}=1$. 
	
	\subsubsection{Consensus stage cost} We simple select as consensus stage cost $l_c(\eta,v_\eta)=\eta'Q_c\eta+v_\eta'U_c v_\eta$, with $Q_c\succ0 $ and $\lambda_{max}(Q_c+\lambda_{\eta}^2 U_c)\leq m_\eta\lambda_\eta$, which clearly satisfies Assumption~\ref{sc}.\ref{asciii}. In this example we use $\lambda_\eta=Q_c=U_c=1$ and $m_\eta = \lambda_{max}(Q_c+\lambda_{\eta}^2 U_c)/\lambda_\eta =2$.
	\begin{figure}[t!]
		\begin{center}
			\includegraphics[width=7cm]{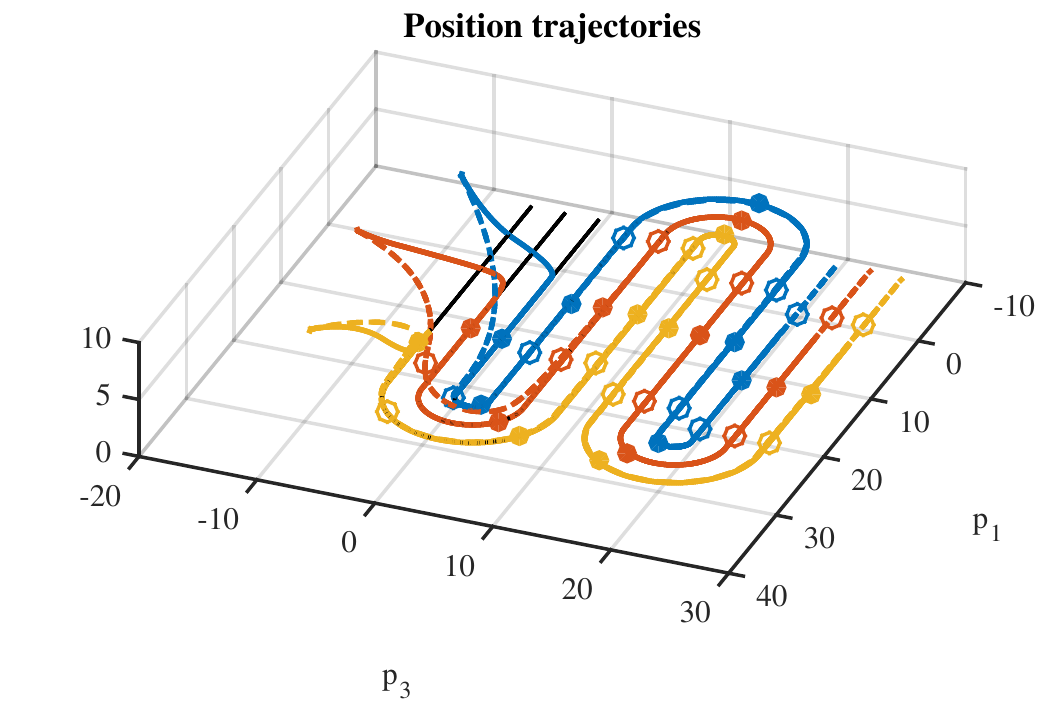}
			\vspace{-0.4cm}
			\caption{The black lines denote the desired paths $c_d^{[i]}(t)$, with $i=1,2,3$; the yellow, red, and blue lines denote the closed-loop trajectories of $c^{[i]}(t)$, with $i=1,2,3$, where the dashed lines are associated to the case $Q=0.1I_{3\times3}$ and the solid lines refer to the case $Q=100I_{3\times3}$. The solid and empty dots denote the positions of the vehicles every 10 seconds associated with the solid and dashed lines, respectively. }
			\label{fig:pos}
		\end{center}
		\begin{center}
			\includegraphics[width=7cm]{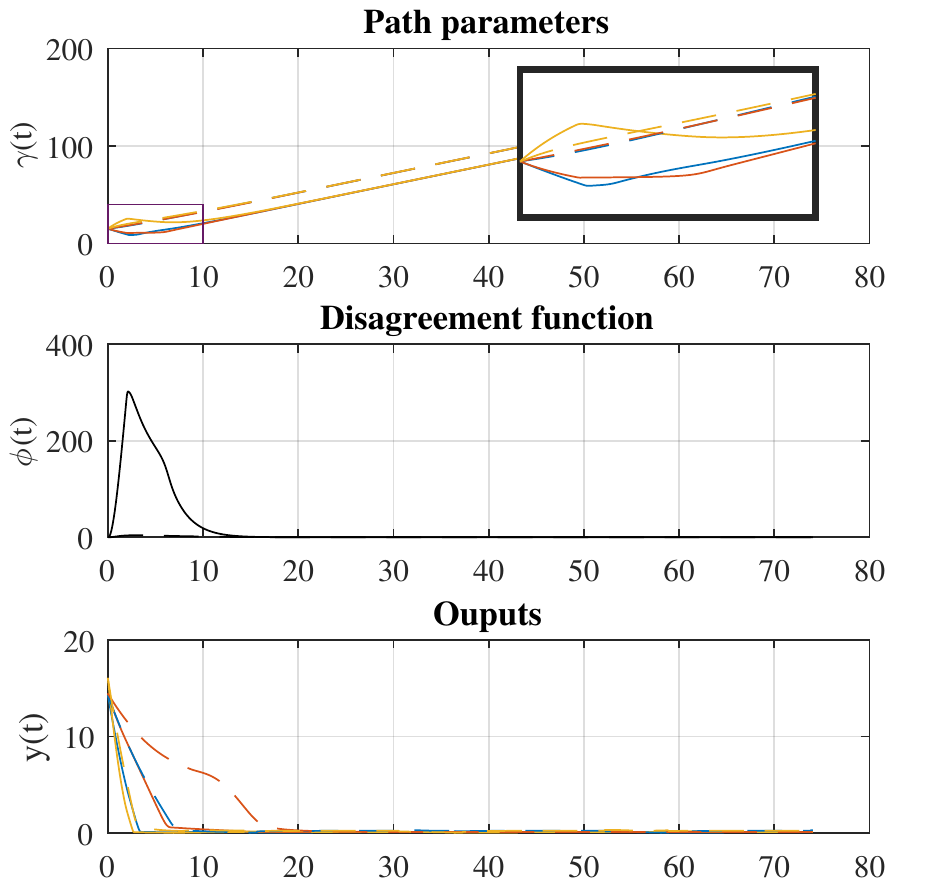}
			\vspace{-0.3cm}
			\caption{The dashed lines are associated to the case $Q=0.1I_{3\times3}$ and the solid lines refer to the case $Q=100I_{3\times3}$.\vspace{-0.5cm}}
			\label{fig:pos2}
		\end{center}
		\vspace{-0.3cm}
	\end{figure}

	\vspace{-0.3cm}
	\subsection{Numerical simulations}\label{sec:nres}
	
	The MPC controller is designed with horizon length $T\!\!=\!0.4$ $s$. The constraint \eqref{eq:ncon} is never active. At time $t_0$ the internal state of the controller is initialized with $\eta_0^{[i]}=0$ and $\gamma_0^{[i]}=15$ for all $i\in\mathcal{I}$, i.e., at consensus. Two simulation scenarios are considered. First, selecting a strong weight on the tracking error, i.e., choosing $Q = 100 I_{3\times3}$, and then reducing such weight to $Q=0.1 I_{3\times3}$ and therefore favoring the consensus. Fig.~\ref{fig:pos} displays the associated closed-loop state trajectories and Fig.~\ref{fig:pos2} the time evolution of the path parameter, disagreement function, and output signals. 
	
	The simulations show that the proposed scheme, depending on the choice of the parameter $Q$, optimally exploits the coordination parameters in order to increase the convergence rate of the output signals. For the case with $Q=100I_{3\times 3}$ the controller initially drives the coordination parameters out of consensus to favor the reduction of the path error. Then, once the output is decreased, the controller drives the coordination vectors back to consensus. In the case $Q=0.1 I_{3\times 3}$, we notice the same behavior, but where the value of the disagreement function is kept almost zero. This latter case resembles a decoupled, and undesired, design where the consensus of the coordination parameters is carried out independently from the output tracking problem.
	
	\subsubsection{Decoupled approach}
\begin{figure}[t!]
		\begin{center}
			\includegraphics[width=7cm]{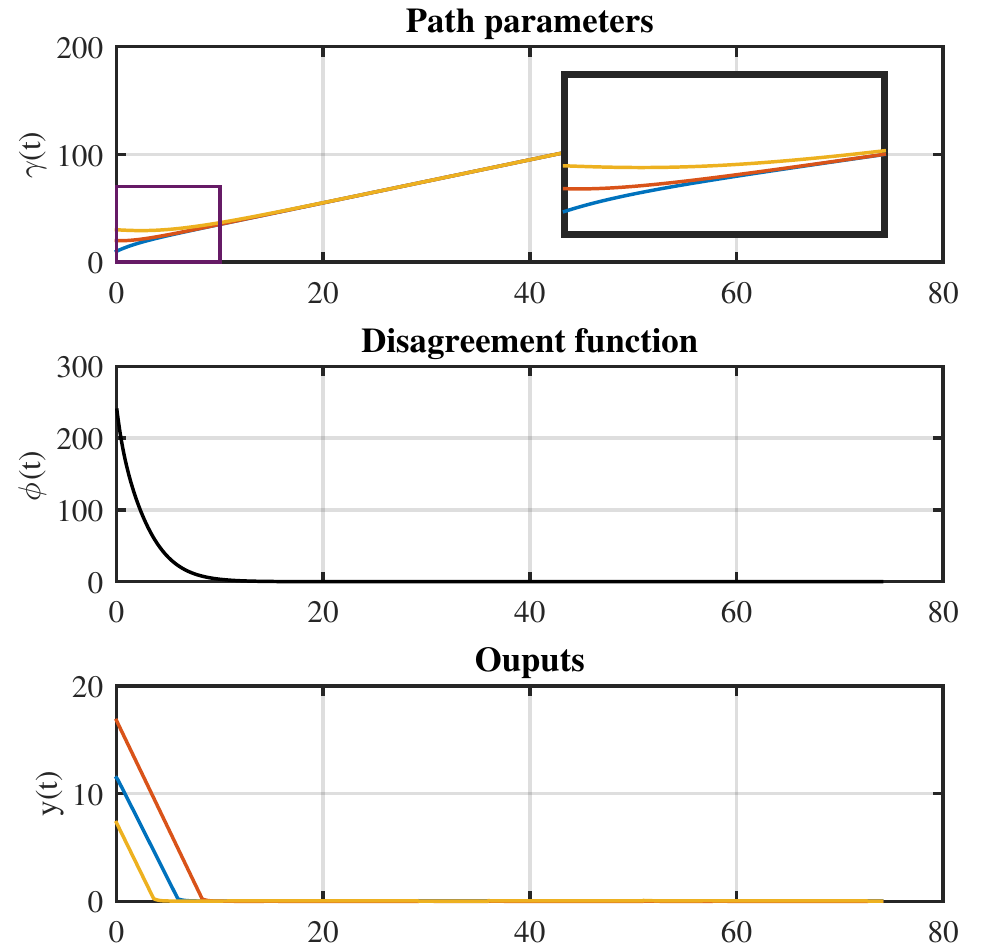}
			\vspace{-0.3cm}
			\caption{Closed-loop signal associated with the simulation of the decoupled approach.\vspace{-0.5cm}}
			\label{fig:ex2pos2}
		\end{center}
		\vspace{-0.5cm}
	\end{figure}

	In contrast to the proposed procedure, we now compare the numerical results with a simpler decoupled strategy that can be found in the literature (e.g., \cite{xiang2012synchronized}). This decoupled approach consists in designing a first controller that drives the vehicle to its path, for any velocity of the coordination vector, and a second controller that drives the coordination vectors to consensus and to a common velocity. Connecting these two controllers we obtain a solution to Problem \ref{pm:coodrinated-output}. For instance, we could achieve this scheme using the path following control law \eqref{eq:exAux} and the consensus law in Proposition \ref{lem:assat}, respectively. Although this strategy is appealing due to its simplicity, it presents some strong limitations. Specifically, since there is no feedback from the position of the vehicle to the consensus law, the vehicle will never be able to adapt their speed to help each others by retarding or accelerating their motion to favorite the decrease of the output signal (tracking error) at the expenses of coordination. Indeed, one can see in Fig. \ref{fig:ex2pos2} the simulations results with the simplified version, where it is clear by comparing with Fig. \ref{fig:pos2} (solid line) that the evolution of the coordination vector evolves independently from the tracking error and follows a standard exponential convergence to consensus. This is due to the absence of feedback with the real position of the vehicle. In contrast, in Fig. \ref{fig:pos} and Fig. \ref{fig:pos2} (solid line, i.e., when the tracking error is highly penalized in the performance index) the coordination vectors adjust and move out of consensus to favorite the decreased of the tracking error. Specifically, in Fig. \ref{fig:pos} (solid line) the vehicles first move to their paths and then are driven to the formation. This is in contrast to the decoupled approach where the vehicles are forced to follow the point associated with the coordination vectors that move independently from their position, therefore delaying the convergence to zero of the tracking error.  Moreover, it is important to stress that the decoupled approach suffers from the fact that the satisfaction of the input constraints can be challenging to verify/satisfy and might lead to inadequate performances (e.g., by implying extremely low gains of the controllers in order to have input signals with small magnitude).
		 Adopting the proposed framework, we are able to optimally balance the output regulation objective with the consensus goal while enforcing hard constraints.
	
\vspace{-0.2cm}
	\section{Conclusions} \label{sec:Conclusions}
	
	The paper presents a distributed MPC scheme for coordinated output regulation with application to the CPF problem of multiple robotic vehicles. The proposed solution is non-iterative and distributed, in the sense that the exchange of information is only performed among neighboring vehicles and only take place once every $t\in\mathcal{T}$. In contrast to similar approaches described in the literature, the proposed control framework optimally balances the output regulation goal with the objective of driving the coordination parameters to consensus, achieving the coordination of the vehicles without using of consistency-like constraints. 
	The simulation results show that being able to optimally drive the coordination parameter can lead to a significant increase of convergence rate of the output signals, which is desired in many practical applications. Note however that since the presented scheme is optimization based, the computation burden required to compute the input might become a limitation in case of highly nonlinear systems or long prediction horizon. As a future direction, we wish to validate the proposed approach on the field with practical experiments.
	
	\appendices
	\section{Proof of Theorem \ref{th:main}} \label{app:main}
	
	The proof is structured as follows. First, the value function used for the analysis is introduced. Then, we define and construct the shifted trajectories that are used (i) to prove the recursive feasibility of the open-loop MPC problem and (ii) to analyze the evolution of the value function, where in the latter step we also consider the effect of the interaction with the other vehicles in the network. The proof is concluded with the convergence result obtained using Barbalat's lemma.
	
	\par {\bf Value function definition.} Consider the value function $
	V(t):=J_T^*(p(t))$ which is obtained by solving the problem $\mathcal{P}(p(t))$ with $
	p( t )=(t,x(t),\gamma(t),\gamma_{\mathcal{N}}(t),\eta(t)). $ Notice that, although the implementation of \eqref{eq:kmpc} only requires to solve $\mathcal{P}(p(t))$ at the time instants $t\in\mathcal{T}$, the value function considers its optimal value at every time $t\geq t_{0}$, and therefore is well defined for all time instants.
	
	\par {\bf Shifted trajectories design.} We recall that, for a generic time $t\geq t_0$, we use the superscript $^{\star t}$ to denote all the trajectories of a given signal associated with the optimal predictions of $\mathcal{P}(t)$. Next, the optimizers of the problem $\mathcal{P}(t_k)$ are used to build the shifted inputs $\bbaru^{s_{t_k+\delta}}\in\mathcal{PC}(t_k+\delta, t_k+\delta+T)$ and $\bbarv_\gamma^{s_{t_k+\delta}}\in\mathcal{PC}(t_k+\delta, t_k+\delta+T)$, with $\delta\in(0,T)$, as follows
	\begin{subequations}\label{eq:us}
		\vspace{-1mm}\begin{align}
		&\bar{u}^{s_{t_k+\delta}}(\tau)~\!\!\!:= \!
		\begin{cases}
		\bar{u}^{\star t_k}(\tau),& \!\!\!\!\tau \!\in\! [t_k\!+\!\delta,t_k\!+\!T), \\
		k_{aux}(\tau,x,\gamma,u_{\gamma}), &  \!\!\!\!\tau \!\in \![t_k\!\!+\!T\!,t_k\!\!+\!T\!+\!\delta],
		\end{cases}\\
		&\bar{v}_\gamma^{s_{t_k+\delta}}(\tau)~\!\!\!:=~\!\!\!\begin{cases}
		\bar{v}^{\star t_k}(\tau),& \tau \in [t_k+\delta,t_k+T),  \\
		- \lambda_\eta \bar{\eta}(\tau) , & \tau \in  [t_k+T\!,t_k+T+\delta].
		\end{cases}\\
		&\bar{u}^{s_{t_k+\delta}}_{\gamma}(\tau)\!:=\bar{u}_{\gamma,aux_{t_k+\delta}}(\tau) + \bar{\eta}^{s_{t_k+\delta}}(\tau)
		\end{align}\\[-0.3cm]
	\end{subequations}
	with $\bar{u}_{\gamma,aux_{t_k+\delta}}(\cdot)$ from \eqref{eq:ubaric}, and where the superscript $^{s_{t_k+\delta}}$ is used to refer a \emph{shifted trajectory}, meaning an open-loop trajectory associated to $J_T(p(t_k+\delta),\bbaru^{s_{t_k+\delta}},\bbarv_\gamma^{s_{t_k+\delta}})$.
	
	\par {\bf Recursive feasibility.} Next, we show that the shifted trajectories are feasible for $J_T(p(t_k+\delta),\bbaru^{s_{t_k+\delta}},\bbarv_\gamma^{s_{t_k+\delta}})$.
	
	\underline{Analysis of $(\bbaru,\bbarx,\bbarv_\gamma,\bbare)$}: For all $\tau\in[t_k+\delta,t_k+T)$ the shifted trajectories $(\bbaru^{s_{t_k+\delta}},\bbarx^{s_{t_k+\delta}},\bbarv_\gamma^{s_{t_k+\delta}},\bbare^{s_{t_k+\delta}})$ are feasible since, in that interval, they corresponds to trajectories $(\bbaru^{\star t_k},\bbarx^{\star t_k},\bbarv_\gamma^{\star t_k},\bbare^{\star t_k})$, which were feasible at time $t_k$. 
	For all $\tau\in[t_k+T,t_k+T+\delta]$ we have
$
		\bar v_\gamma^{s_{t_k+\delta}}(\tau)=-\lambda_\eta\bar\eta(\tau)  
		\implies\bar\eta(\tau)=\bar\eta(t_k+T)e^{-\lambda_\eta (\tau-t_k-T)}  
		\implies \|\bar\eta(\tau)\|\leq\|\bar\eta(t_k+T)\|\leq r_\eta 
		\implies \|\bar v_\gamma^{s_{t_k+\delta}}(\tau)\|\leq  \lambda_\eta r_\eta.$
	Combining this result with  Assumption~\ref{sc2}.\ref{item:sc11}, it follows that the input constraint $\bar v_\gamma(\tau)\in\mathcal{V}_\gamma(\tau)$ and the state constraint \eqref{eq:ncon} are satisfied. Moreover,  $(\bbaru^{s_{t_k+\delta}},\bbarx^{s_{t_k+\delta}})$ are kept feasible by the auxiliary regulation controller from Assumption~\ref{sc}.\ref{item:sc1aux}.
	\\
	\underline{Analysis of $(\bbarg,\bbaru_\gamma)$}: The variables $(\bbarg^{s_{t_k+\delta}},\bbaru_\gamma^{s_{t_k+\delta}})$ are not constrained, and therefore are trivially feasible. 
	
	\par {\bf Unknown future behavior of the neighboring systems.} 
	It is important to notice that, in contrast to $(\bbaru,\bbarx,\bbarv_\gamma,\bbare)$, the variables $(\bbarg^{s_{t_k+\delta}},\bbaru_\gamma^{s_{t_k+\delta}})$ do not match $(\bbarg^{\star t_k},\bbaru_\gamma^{\star t_k})$ in the common time interval $[t_k+\delta,t_k+T)$. Specifically, from the definition in \eqref{eq:ubaric}, $\bar{u}_{\gamma,aux_{t_k+\delta}}(\tau)$ differs from $\bar{u}_{\gamma,aux_{t_k}}(\tau)$ in the interval $\tau\geq t_k+\delta_k$. For the case $\delta = \delta_k$ this difference captures the discrepancy between the predictions obtained using the network measurements at time $t_k$ and the one obtained using the measurements at time $t_{k+1}$. The effect of this distinction on the convergence analysis is considered in Lemma \ref{lem:deltaaux}, which will be called in the next developments. 
	
	\par {\bf Value function decrease.} 
	Before analyzing the whole value function, notice that along the shifted trajectories with $\tau \in [t_k+T,t_k+T+\delta]$, i.e., when $\bar{v}_\gamma^{s_{t_k+\delta}}(\tau)=-\lambda_\eta \bar\eta^{s_{t_k+\delta}}(\tau)$, integrating \eqref{eq:inmcon} from $t_k+T$ to $t_k+T+\delta$ results in 
	\vspace{-1mm}\begin{align}
	&{1\over 2}m_\eta\left(\bar{\eta}^{s_{t_k+\delta}}(t_k+T+\delta)\right)^2-{1\over 2}m_\eta\left(\bar{\eta}^{s_{t_k+\delta}}(t_k+T)\right)^2 \nonumber \\
	&\qquad\leq -\int_{t_k+T}^{t_k+T+\delta}\bar{l}_c^{s_{t_k+\delta}}(\tau)d\tau \label{eq:mdiff}
	\end{align} \\[-0.3cm]
	where to simplify the notation we use $\bar l_c(\tau)$ to denote $l_c(\cdot)$ evaluated at the predicted signals as function of $\tau$.  
	Similarly, evaluating \eqref{eq:decrese} along the shifted trajectories  $(\bbaru^{s_{t_k+\delta}},\bbarx^{s_{t_k+\delta}},\bbarv_\gamma^{s_{t_k+\delta}},\bbare^{s_{t_k+\delta}})$ we have
	\vspace{-1mm}\begin{align}
	\bar{m}^{s_{t_k\!+\!\delta}}(t_k\!+\!\delta\!+\!T) \!+\!\!\!\int_{t_k+T}^{t_k+T+\delta}\!\!\!\!\!\!\!\!\!\!\!\!\bar{l}^{s_{t_k+\delta}}(\tau)d\tau\! \leq\!   \bar{m}^{s_{t_k+\delta}}(t_k\!+\!T) . \label{eq:mdiff2}
	\end{align} \\[-0.3cm]
	The shifted trajectories are now used in combination with the optimality of the MPC controller to analyze the evolution of the value function. Consider first $\delta\in[0,t_{k+1}-t_k]$, for the generic $t_k\in\mathcal{T}$, we have
{\small
\vspace{-1mm}\begin{align*}
&V(t_k+\delta) \leq J_T(p(t_k+\delta),\bbaru^{s_{t_k+\delta}},\bbarv_\gamma^{s_{t_k+\delta}}) \\
&= V(t_k)\!-\! \int_{t_k}^{t_k+\delta}\!\bar l(\tau)\! +\! \bar l_c(\tau)d\tau 
	+ \int_{t_k+\delta}^{t_k+T} 
	\!\bar l^{s_{t_k+\delta}}(\tau)\!-\! \bar{l}^{\star t_k}(\tau)d\tau\\
&\qquad	+ \bar{m}^{s_{t_k+\delta}}(t_k+\delta+T) -  \bar{m}^{\star t_k}(t_k+T) \\
&\qquad	+ \int_{t_k+T}^{t_k+T+\delta}\bar{l}^{s_{t_k+\delta}}(\tau) + \bar{l}^{s_{t_k+\delta}}_c(\tau)d\tau  \\
&\qquad + {1\over 2}m_\eta\big(\bar{\eta}^{s_{t_k+\delta}}(t_k+T+\delta)\big)^2 	-  {1\over 2}m_\eta\big(\bar{\eta}^{s_{t_k+\delta}}(t_k+T)\big)^2  \\
&	\leq V(t_k)\!-\! \int_{t_k}^{t_k+\delta}\!\bar l(\tau)\! +\! \bar l_c(\tau)d\tau
	+ \int_{t_k+\delta}^{t_k+T} 
	\!\bar l^{s_{t_k+\delta}}(\tau)\!-\! \bar{l}^{\star t_k}(\tau)d\tau \\
&\qquad	+ \bar{m}^{s_{t_k+\delta}}(t_k+T) -  \bar{m}^{\star t_k}(t_k+T)\\
&	\leq V(t_k)- \int_{t_k}^{t_k+\delta}\bar l(\tau) + \bar l_c(\tau)d\tau +\bar{\beta}_{\delta}(t_k) 
\end{align*}} \\[-0.3cm]
	for an integrable function $\bar{\beta}_{\delta}(\cdot)$, where the first inequality stems from the sub-optimality of the shifted variables $(\bbaru^{s_{t_k+\delta}},\bbarv_\gamma^{s_{t_k+\delta}})$, the second equality uses the fact that $\bar l^{s_{t_k+\delta}}(\tau) - \bar{l}_c^{\star t_k}(\tau) = 0$ for $\tau\in[t_k+\delta, t_k+T)$,
	the second inequality is due to \eqref{eq:mdiff} and \eqref{eq:mdiff2}, and the last one is obtained from Lemma~\ref{lem:deltaaux} in the Appendix \ref{app:lemmas}. Since the latter inequity holds on any interval $[t_k,t_k+\delta]$ with $\delta \!\in\! [0,t_{k+1}\!\!-\!t_k]$, we can analyze the generic case $\delta\!>\!0$ as follows
	\vspace{-1mm}\begin{align}
	&V(t_k+\delta)=V(t_k)+\!\!\!\!\!\!\!\!\sum_{\substack{t_j\in\mathcal{T}\\t_k\leq t_j<\lfloor t_k+\delta \rfloor}}\{V(t_{j+1})-V(t_{j})\}\nonumber\\
	& \!+\!V(t_{k}+\delta)\!-\!V(\lfloor t_k+\delta \rfloor)\!\leq\! V(t_k)\!-\! \int_{t_k}^{t_k+\delta}\!\!\!\!\!\!\!\!\alpha(\|z(\tau)\|)d\tau  + \bar{\bar{\beta}}\label{eq:decVV} 
	\end{align} \\[-0.4cm]
	where we used the fact that the summation $\bar{\bar{\beta}}:=\bar{\beta}_{t_k+\delta-\lfloor t_k+\delta \rfloor}(\lfloor t_k+\delta \rfloor)+\sum_{\substack{t_j\in\mathcal{T}\\t_k\leq t_j<\lfloor t_k+\delta \rfloor}}\bar{\beta}_{\delta}(t_j) $  converges from Lemma \ref{lem:deltaaux}, and the bound $ \bar l(\tau) + \bar l_c(\tau)\geq\alpha(\|z(\tau)\|) $ from Assumption~\ref{sc}.\ref{ascii} and Assumption~\ref{sc2}.\ref{asciii} 
	where $z(t):=[y(t)',\eta(t)']'$ and $\alpha:\mathbb{R}_{\geq 0}\to \mathbb{R}_{\geq 0}$ is a class-$\mathcal{K}_\infty$ function such that $\alpha(\|z\|)\leq \alpha_s(\|y\|)+\alpha_c(\|\eta\|)$. By Lemma~\ref{lem:ydot} in the Appendix \ref{app:lemmas}, the derivative of $\by$ is bounded for bounded values of $\by$, and the same apply to the signal $\be$, from  $\mathcal{V}_\gamma(\cdot)$ being uniformly bounded over time by Assumption~\ref{sc2}.\ref{item:sc11}, and therefore also the derivative of the signal $\bz$ is bounded for bounded values of $z$.
	Then, applying Lemma 37 of \cite{Alessandretti2016} on  $\bz$ we have that for any $t\geq t_0$ and a given $\epsilon>0$ there exists a class-$\mathcal{K}_\infty$ functions $\alpha_{\epsilon}:\mathbb{R}_{\geq 0}\to \mathbb{R}_{\geq 0}$, possibly dependent on $\epsilon$, such that $\int_{t}^{t+\epsilon}\alpha(\|z(\tau)\|)d\tau\geq\alpha_{\epsilon}(\|z(t)\|).$
	As a consequence, using the definition \eqref{eq:defJ} of $J_T(\cdot)$, for any $\delta_1\in(0,T]$ 
	we have that
	$V(t)\geq \alpha_{\delta_1}(\|z(t)\|)$ 
	that is obtained choosing $\epsilon=\delta_1$.
	Combining the latter with the fact that, from \eqref{eq:decVV}, $V(t_0+\delta)\leq V(t_0)+\bar{\bar{\beta}}$ for all $\delta \geq 0$, implies that the trajectory $\bz$ is bounded within the set \vspace{-1mm}\begin{align}
	\left\{ z\in\mathbb{R}^{p+n_c}:\alpha_{\delta_1}(\|z\|)\leq  V(t_0)+\bar{\bar{\beta}}\right\} \label{eq:iset}
	\end{align}\\[-0.4cm]  
	for all $t\geq t_0$, where the set is compact because it is a level set of a class-$\mathcal{K}_\infty$ function, and therefore also the state $x(t)$ of the system is uniformly bounded over time by \eqref{eq:issy} and the boundedness of the input in Assumption~\ref{sc}.\ref{item:sc1}.
	
	At this point, in order to prove convergence of $z(t)$ to the origin as $t\to\infty$, we can use Barbalat's lemma (see, e.g., Lemma 8.2 in \cite{khalil2002nonlinear}). Toward that, next we show that $\alpha(\|z(\tau)\|)$ is a uniformly continuous function of $\tau$. 
	
	First notice that $y(\tau)$ is uniformly continuous on $\tau$ since, by Lemma~\ref{lem:ydot}, possesses a uniformly bounded time derivative for bounded values of $y$, where $y$ is uniformly bounded from $z$ being uniformly bounded in \eqref{eq:iset}. Similarly, $\eta(\tau)$ is uniformly continuous on $\tau$, from  $\mathcal{V}_\gamma(\cdot)$ being uniformly bounded over time. As a consequence, the combined vector $z(\tau)$ is uniformly continuous on $\tau$. Then, we notice that the function $\alpha(\|z(\tau)\|)$  is a uniformly continuous function of $\tau$ because (i) the function $\alpha(\|z(\tau)\|)$ is continuous in $z$ with $z$ bounded within \eqref{eq:iset} and (ii) $z$ is uniformly continuous in $\tau$. 
	
	At this point, the inequality \eqref{eq:decVV} implies $\lim_{\delta \rightarrow \infty}\int_{t_0}^{t_0+\delta}\alpha(\|z(\tau)\|)d\tau~\leq V(t_0)+\bar{\bar{\beta}}<\infty$ where the limit exists since the function $\int_{t_0}^{t_0+\delta} \alpha(\|z(\tau)\|) d\tau $ is non decreasing in $\delta$, from $\alpha(\cdot)$ being non negative, and it is upper bounded by $V(t_0)+\bar{\bar{\beta}}$. Thus, by Barbalat's lemma, $\alpha(\|z(t)\|)\to 0$ as $t\to \infty$ and, by the continuity and positive-definitiveness of $\alpha(\cdot)$, the vector $z(t)$, and therefore the vectors $y(t)$ and $\eta(t)$, converges to the origin as $t\to \infty$. The proof is concluded by noticing that, by Assumption~\ref{ass:con}, as $\eta(t) \to 0$ also the disagreement function is driven to zero.\hfill $\blacksquare$
		\vspace{-4mm}
	\subsection{Lemmas} \label{app:lemmas}
	
	\begin{lem}\label{lem:deltaaux}
		Consider the open-loop MPC problem from Definition~\ref{eq:P} and let Assumption~\ref{ass:con} hold. Let the superscripts $^{s_{t_k+\delta}}$ and $^{\star t_k}$ denote all the trajectories associated with the shifted inputs \eqref{eq:us} and the optimal solutions of $\mathcal{P}(t_k)$, respectively. Then, for any $t_k\!\in\!\mathcal{T}$ and $\delta\!\in\![0,t_{k+1}\!-\!t_k]$ the following holds $\int_{t_k+\delta}^{t_k+T}\!\bar{l}^{s_{t_k+\delta}}(\tau)\!-\!\bar{l}^{\star t_k}(\tau)d\tau\!+\!\bar{m}^{s_{t_k+\delta}}(t_k+T)\!-\!\bar{m}^{\star t_k}(t_k\!+\!T) \!\leq\!\bar{\beta}_{\delta}(t_k) $ for an integrable non-increasing scalar function $\bar{\beta}:\mathbb{R}_{\geq t_0}\to \mathbb{R}_{\geq 0}$ with $
		\int_{t_0}^{+\infty} \bar{\beta}_{\delta}(\tau)d\tau \leq b_{\bar{\beta}} < \infty$
		and constant $b_{\bar{\beta}}\in\mathbb{R}_{\geq0}$.
	\end{lem}
	\begin{IEEEproof}
		From the definitions of $\bar{u}_{\gamma}(\cdot)$ in \eqref{eq:ubari} and $\bar{u}_{\gamma,aux_{t}}(\cdot)$ in \eqref{eq:ubaric}, for all $\tau\in[t_k+\delta,t_k+T)$ we have that 
		\begin{subequations}\label{eq:dds}
			\vspace{-1mm}\begin{align}
			\bar{u}^{\star t_k}_{\gamma}(\tau) &=\bar{u}_{\gamma,aux_{t_k}}(\tau) + \bar{\eta}^{\star t_k}(\tau)\\
			\bar{u}^{s_{t_k+\delta}}_{\gamma}(\tau) &=\bar{u}_{\gamma,aux_{t_k+\delta}}(\tau) + \bar{\eta}^{\star t_k}(\tau).
			\end{align} \\[-0.4cm] 
		\end{subequations}
		The constraint \eqref{eq:ncon} enforces \eqref{eq:boundeta}, and therefore by Assumption~\ref{ass:con} we have $
		\| \bar{u}^{\star t_k}_{\gamma}(\tau) -\bar{u}^{s_{t_k+\delta}}_{\gamma}(\tau) \| =\|\bar{u}_{\gamma,aux_{t_k}}(\tau) -\bar{u}_{\gamma,aux_{t_k+\delta}}(\tau)\|
		 \leq 2\left\|{1\over \delta_k}k_{con}(\gamma^{[i]}(t_k),\gamma_{\mathcal{N}^{[i]}}(t_k)) \right\|
		\leq {2\over \delta_{lb}}\beta(\sum_{i\in\mathcal{I}}\|\gamma_0^{[i]}\|,k)$
		where $k$ is the subscript of $t_k$. Combining the latter with \eqref{eq:dotgamma} we also obtain that
		$
		\| \bar{\gamma}^{\star t_k}(\tau) -\bar{\gamma}_{aux_{t_k}}(\tau) \| \leq {2T\over \delta_{lb}}\beta(\sum_{i\in\mathcal{I}}\|\gamma_0^{[i]}\|,k)\nonumber
		$
		where  $\tau\in[t_k,t_k+T)$ and where the last inequality comes from $\tau$ being less than $t_k+T$. At this point, combining the latter inequalities with the Lipschitz properties of Assumption~\ref{sc3} we can write
		$
		\int_{t_k+\delta}^{t_k+T}l(\tau,\bar{x}^{\star t_k},\bar{u}^{\star t_k},\bar{\gamma}_{aux_{t_k}},\bar{u}_{\gamma, aux_{t_k}}) - \bar{l}^{\star t_k}(\tau)d\tau \leq{\footnotesize  \int_{t_k+\delta}^{t_k+T} {C_l} \left\|\begin{bmatrix}\bar{\gamma}_{aux_{t_k}}-\bar{\gamma}^{\star t_k}\\\bar{u}_{\gamma,aux_{t_k}}-\bar{u}^{\star t_k}_{\gamma}\end{bmatrix}\right\| }
		\leq \tfrac{2C_lT(1+T)}{\delta_{lb}}\beta(\sum_{i\in\mathcal{I}}\|\gamma_0^{[i]}\|,k)
		$
		and, similarly, $
		m(t_k+\delta+T,\bar{x}^{s_{t_k}},\bar{\gamma}_{aux_{t_k}}) -  m(t_k+T,\bar{x}^{\star t_k},\bar{\gamma}^{\star t_k}_i) \leq C_m \left\|\begin{bmatrix}\bar{\gamma}_{aux_{t_k}}-\bar{\gamma}^{\star t_k}\\\bar{u}_{\gamma,aux_{t_k}}-\bar{u}^{\star t_k}_{\gamma}\end{bmatrix}\right\| \leq  \tfrac{2C_m(1+T)}{\delta_{lb}}\beta(\sum_{i\in\mathcal{I}}\|\gamma_0^{[i]}\|,k).$
		Therefore, the lemma holds with  $\bar{\beta}_{\delta}(t)\!=\!{2(C_lT+C_m)(1+T)\over \delta_{lb}}\beta(\sum_{i\in\mathcal{I}}\|\gamma_0^{[i]}\|,k)$, $t\in[t_k,t_{k+1})$, which is an integrable non-increasing function because $\beta(\cdot)$ is assumed to be integrable and non-increasing. 
	\end{IEEEproof}
	
	\begin{lem} \label{lem:ydot}
		Consider the system \eqref{eq:allsys}, let Assumption~\ref{ass:1}-\ref{ass:y} and~\ref{ass:con} hold and let $\mathcal{V}_\gamma(\cdot)$ and $\mathcal{U}(\cdot)$ be uniformly bounded over time. Then, the time derivative of $y(t)$ is uniformly bounded over time for any trajectory of $y(t)$ uniformly bounded over time. 
	\end{lem}
	
	\begin{IEEEproof}
		From the assumptions stated in this lemma, one can conclude that there exist scalars $b_y\geq 0$, $b_u\geq 0$, and $b_g\geq 0$ such that $\|y(t)\|\leq b_y$,  $\|u(t)\|\leq b_u$, and $\|g(t)\|\leq b_g$ uniformly over time. Then, from \eqref{eq:issy}, the evolution of the state vector is uniformly bounded over time as $\|x(t)\|\leq B:=\beta_x(\|x_0\|,0)+\sigma_y(b_y)+\sigma_u(b_u)+\sigma_x$ and therefore, by Assumption~\ref{ass:1}, also its derivative is bounded as $\|f(\cdot)\|\leq b_f$ for some scalar $b_f\geq0$. Moreover, we recall that in Assumption~\ref{ass:con}, the consensus controller provides bounded inputs, i.e.,$\|k_{con}(\cdot)\|\leq b_c$ for some constants $b_c\geq 0$. Combining the latter facts with Assumption~\ref{ass:y}.\ref{item:yDotB}, it follows that the time derivative of $y(t)$, given by
		$\dot{y}(t)\!=\!\nabla h(t,x,\gamma)\begin{bmatrix}1 &f(t,x,u) &{g}(t)+u_\gamma\end{bmatrix}'$ 
		is bounded as $\|\dot{y}(t)\|\!\leq\! b_B(1\!+\!\|f(t,x,u)\|\!+\!b_g\!+\!\|u_\gamma\|)\!\leq\!b_B(1\!+\!b_f\!+\!b_g\!+\!b_c\!+\!a_\eta e^{-\lambda_\eta t_0})$ where we used the fact that $u_\gamma$ is chosen as in \eqref{eq:ubari} where $\bar{\eta}$ is constrained by \eqref{eq:ncon}.  
	\end{IEEEproof}

	\section{Analysis of the consensus control law}\label{app:analycon}
	Consider a network of discrete-time systems as stated in Assumption~\ref{ass:con}. This section shows that the closed-loop system \eqref{eq:conSys} with \eqref{eq:conscon} satisfies Assumption~\ref{ass:con}. For the sake of simplicity, we consider the scalar case, i.e., $n_c=1$, which is the case used in the Section \ref{sec:numerical} for the numerical results.
	\vspace{-0.4cm}
	\subsection{Background} 
	We start by recalling some results from \cite{Olfati-Saber2007,consensus02} that are used in the analysis of the consensus control law \eqref{eq:conscon}.
	
	The matrix $A$ denotes the \emph{adjacency matrix} of the graph $\mathcal{G}$, where the generic component at row $i$ and column $j$ is defined as $[A]_{i,j}=a_{ij}$, for positive scalars $a_{ij}>0$, if $(i,j)\in\mathcal{E}$ and $[A]_{i,j}=0$, otherwise. The associated \emph{Laplacian} is defined as $L=D-A$ where  $D=\diag(d_1,\dots,d_{|\mathcal{V}|})$ denotes the \emph{degree matrix} with generic diagonal component $d_i=\sum_{j\neq i}a_{ij}$. The matrix $P=I_{|\mathcal{V}|}-\epsilon L$ denotes the corresponding \emph{Perron matrix} with parameter $\epsilon$.  A nonnegative matrix is called row (column) \emph{stochastic} if its row-sums (column-sums) are one. A \emph{doubly stochastic} matrix is a matrix that is row stochastic and column stochastic. A matrix is called \emph{primitive} if it has only one eigenvalue with maximum modulus, and it is called \emph{irreducible} if its associated graph is strongly connected.
	
	\begin{lem}[Lemma 3 of \cite{Olfati-Saber2007}]\label{lem:clo} Let $\mathcal{G}$ be a digraph with \emph{maximum degree} $\Delta = \max_{i}(\sum_{j\neq i}a_{ij})$. Then, the Perron matrix P with parameter $\epsilon \in (0,1/\Delta]$ satisfies the following properties: 1) P is a row stochastic nonnegative matrix with a trivial eigenvalue of $1$; 2) All eigenvalues of P are in the unit circle; 3) If $\mathcal{G}$ is a balanced graph, then P is a doubly stochastic matrix; 4) If $\mathcal{G}$ is strongly connected and $\epsilon\in (0,1/\Delta)$, then P is a primitive matrix. \hfill $\square$
	\end{lem}
	
	The \emph{disagreement vector} of the network is defined as 
	\vspace{-1mm}\begin{align}
	\delta(k) := \xig(k) - \alpha(k) \one,&& \alpha(k) := {1\over n}\one'\xig(k) \label{eq:disagreement2}
	\end{align}\\[-0.3cm]
	with $\xig:=(\xi^{[1]},\xi^{[2]},\dots, \xi^{[|\mathcal{V}|]})'$ and $n=|\mathcal{V}|$.
	
	\begin{thm}[Theorem 3 of \cite{Olfati-Saber2007}]\label{thm:cpb} Let $\mathcal{G}$ be a balanced digraph (or undirected graph) with Laplacian L with a symmetric part $L_s=(L+L')/2$ and Perron matrix P. Then, 1)  $\lambda_2=\min_{\one'\delta=0}(\delta'L\delta/\delta'\delta)$ with $\lambda_2=\lambda_2(L_s)$, i.e.,
			$\delta'L\delta \geq \lambda_2\|\delta\|^2$
			for all disagreement vectors $\delta$; 2) $\mu_2=\max_{\one'\delta=0}(\delta'P\delta/\delta'\delta)$ with $\mu_2=1-\epsilon\lambda_2$, i.e.,
			$\delta'P\delta \leq \mu_2\|\delta\|^2$
			for all $\delta$. \hfill $\square$
	\end{thm}
	
	\subsection{Robustness analysis}
	In order to show that \eqref{eq:conscon} satisfies Assumption~\ref{ass:con}, we start by proving that the dynamical model of the disagreement vector \eqref{eq:disagreement2} associated to the closed-loop system \eqref{eq:conSys} with \eqref{eq:conscon} is Input-to-State-Stable (ISS) with respect to $\delta_\eta(k)$ where
	\vspace{-1mm}\begin{align}
	\delta_{\eta}(k):=\etag(k) -\alpha_\eta(k)\one,&&\alpha_\eta(k):={1\over n}\one'\etag(k) \label{eq:def_d_a_eta}
	\end{align}\\[-4mm]
	with $\etag:=(\eta^{[1]},\eta^{[2]},\dots, \eta^{[|\mathcal{V}|]})'$ and $n=|\mathcal{V}|$. That is, we will show that there exists a class-$\mathcal{KL}$ function $\beta:\mathbb{R}_{\geq 0}\times \mathbb{R}_{\geq 0} \to \mathbb{R}_{\geq 0}$ and a class-$\mathcal{K}$ function $\gamma:\mathbb{R}_{\geq 0}\to\mathbb{R}_{\geq 0}$ such that, for all $k\geq k_0$, the inequality
	\vspace{-1mm}\begin{align}
	\|\delta(k)\|\leq \beta(\|\delta(k_0)\|,k-k_0)+\gamma(\|\bd_\eta([k_0,k]))\|_{\infty}) \label{eq:issbeta}
	\end{align}\\[-0.4cm]
	holds, where $\bd_\eta([k_0,k])$ denotes the trajectory $\bd_\eta(\cdot)$ considered over the time steps $k_0, k_0+1,\dots,k$.
	At this point, we refer to \cite[Definition 3.2]{Jiang2001} for a definition of ISS-Lyapunov function. The following result provides an ISS-Lyapunov function that by \cite[Lemma 3.5]{Jiang2001} implies \eqref{eq:issbeta}.
	
	\begin{thm}[ISS property of consensus]\label{thm:isscons}
		Consider the network of discrete-time systems in Assumption~\ref{ass:con} with  $n_c=1$ (scalar state) and \eqref{eq:conscon}, and the disagreement vector of the network defined as \eqref{eq:disagreement2}. If $\mathcal{G}$ is a balanced strongly connected graph and $\epsilon \in (0,1/\Delta)$ where $\Delta = \max_{i}(\sum_{j\neq i}a_{ij})$, then $\Phi(k):=\delta(k)'\delta(k)$ is an ISS-Lyapunov function that satisfies
		\vspace{-1mm}\begin{align}
		\Phi(k+1) \leq (1-\lambda_{\Phi})\Phi(k)+c\| \delta_\eta(k)\|^2 \label{eq:phi_dyn}
		\end{align}\\[-0.4cm]
		with $\delta_\eta(k)$ from \eqref{eq:def_d_a_eta}, for constants $c>0$ and $0<\lambda_{\Phi}<1$. 
	\end{thm}
	\begin{IEEEproof}
		First notice that since $\mathcal{G}$ is a balanced strongly connected graph with $\epsilon \in (0,1/\Delta)$, by Lemma \ref{lem:clo} the Perron matrix $P$ is double stochastic and, therefore, satisfies $\one'P = \one'$ and $P\one~=~\one$. Recalling the definitions \eqref{eq:disagreement2} and \eqref{eq:def_d_a_eta}, the closed-loop \eqref{eq:conSys} with \eqref{eq:conscon} can be rewritten in the compact form
		$\xig(k+1)= P\xig(k)+\etag(k)$, $
		\alpha(k+1)={1\over n}\one'(P\xig(k)+\etag(k))=\alpha(k)+\alpha_\eta(k)$ where we used the fact that $\one'P = \one'$. Moreover, combining the latter with the definition \eqref{eq:disagreement2} of $\delta(k+1)$ and the fact that $P\one~=~\one$, we have $
		\delta(k+1) = P(\delta(k)+\alpha(k)\one)+\etag(k) - \alpha(k)\one-\alpha_\eta(k) \one = P\delta(k) +\delta_{\eta}(k)$
		and therefore$
		\Phi(k+1) = \|P \delta(k)\|^2 + \|\delta_{\eta}(k)\|^2 + 2 \delta_{\eta}(k)'P\delta(k) \leq \mu_2^2\Phi(k) + \|\delta_{\eta}(k)\|^2 + 2 \|\delta_{\eta}(k)\|\|\delta(k)\|$
		where the inequality comes from Theorem \ref{thm:cpb} and the fact that $\|P\|\leq\mu_2\leq 1$. Lastly, we can write $
		\Phi(k+1)-\Phi(k) \leq -0.5(1-\mu_2^2)\Phi(k)
		\quad\quad-0.5(1-\mu_2^2)\Phi(k)+ \|\delta_{\eta}(k)\|^2 + 2 \|\delta_{\eta}(k)\|\|\delta(k)\|
		= -0.5(1-\mu_2^2)\Phi(k) 
		-(a\|\delta(k)\|+b\|\delta_\eta(k)\| )^2+c\| \delta_\eta(k)\|^2 
		\leq  -\lambda_{\Phi}\Phi(k)+c\| \delta_\eta(k)\|^2 
		$
		with $\lambda_{\Phi}=0.5(1-\mu_2^2)<1$, $a = \sqrt{(1-\mu_2^2)/2}$, $b=-\mu_2/a$, $c=b^2+1$, that concludes the proof. 
	\end{IEEEproof}

	\subsection{Proof of Proposition \ref{pro:new}}\label{app:pro:new}
	For the case ${g}(t)=0$, the closed-loop \eqref{eq:gammaDot} with $u^{[i]}_\gamma(t)=\bar{u}^{[i]}_{\gamma,aux_{\lfloor t \rfloor}}(t)$ results in
	$\dot{\gamma}^{[i]}(t)={1\over \delta_k}k_{con}(\gamma^{[i]}(t_k),\gamma_{\mathcal{N}^{[i]}}(t_k))$ 
	for $t\in [t_k,t_{k+1})$, with $k\in\mathbb{Z}_{\geq 0}$. Integrating it, 
	$\gamma^{[i]}(t_{k+1})=\gamma^{[i]}(t_{k}) + k_{con}(\gamma^{[i]}(t_k),\gamma_{\mathcal{N}^{[i]}}(t_k)).$
	This last fact shows that the continuous time evolution of $\gamma(t)$ evaluated at $t=t_k\in\mathcal{T}$ exactly recovers the discrete time evolution of $\xi(k)$ in \eqref{eq:conSys}, with $\eta(k)=0$, and therefore the disagreement function evaluated at $t_k\in\mathcal{T}$ is asymptotically driven to the origin. Moreover, within the generic interval $t\in[t_k,t_{k+1}]$ the state is a linear interpolation of $\gamma^{[i]}(t_{k})$ and $\gamma^{[i]}(t_{k+1})$, i.e., $\gamma^{[i]}(t)\!\!=\!\!\gamma^{[i]}(t_k)\!+\!(t\!\!-\!\!t_k)(\gamma^{[i]}(t_{k+1})\!-\!\gamma^{[i]}(t_{k}))/(t_{k+1}\!\!-\!\!t_k)$ that implies the consensus to be asymptotically reached for the whole continuous-time trajectory. Lastly, since ${g}(t)$ is common to all the vehicles, choosing ${g}(t)\!\neq\!0$ would not affect the value of the disagreement function and, therefore, would still guarantee consensus.\hfill $\blacksquare$
	
	\subsection{Proof of Proposition \ref{lem:assat}}\label{app:plsat}
	
	The proposition is proved by combining the bound \eqref{eq:boundeta} with the ISS property of the consensus law \eqref{eq:conscon} from Theorem  \ref{thm:isscons}. We start by noticing that
	from \eqref{eq:boundeta}, and recalling the definition of $\delta_{\eta}(k)$ in \eqref{eq:def_d_a_eta}, for all $k\geq k_0$ we have $ \|\delta_\eta(k)\|^2=\|\etag(k)\|^2-\|\etag(k)'\one\|^2/|\mathcal{V}|\leq\|\etag(k)\|^2 \leq \sum_{i\in\mathcal{E}} (a^{[i]}_\eta)^2 e^{-2\lambda^{[i]}_\eta (k-k_0)} 
	\leq |\mathcal{V}| (a^{\max}_\eta)^2 e^{-2\lambda^{\min}_\eta (k-k_0)}$
	with $a^{\max}_\eta = \max_{i\in \{1,\dots |\mathcal{V}|\}} a^{[i]}_\eta,\quad \lambda^{\min}_\eta=\min_{i\in \{1,\dots |\mathcal{V}|\}} \lambda^{[i]}_\eta.$ Applying recursively \eqref{eq:phi_dyn} and combining it with the latter inequality results that for a generic $n\geq 0$ we have
	$ \Phi(k_0+n)  \leq (1-\lambda_{\Phi})^n\Phi(k_0) + \sum_{i=0}^{n-1}(1-\lambda_{\Phi})^i c |\mathcal{V}|(a^{\max}_\eta)^2 e^{-2\lambda^{\min}_\eta (n-1-i)} \leq (1-\lambda_{\Phi})^n\Phi(k_0) + d( e^{-2\lambda^{\min}_\eta n} - (1-\lambda_{\Phi})^n )$ with $d = {c |\mathcal{V}|(a^{\max}_\eta)^2 e^{2\lambda^{\min}_\eta}\over 1-(1-\lambda_{\Phi})e^{2\lambda^{\min}_\eta}}$. Since, $\lambda_\Phi<1$ and $\lambda_\eta^{\min}>0$, as $n\to\infty$ the disagreement vector $\|\delta(k_0+n)\|\to 0$ and therefore Assumption~\ref{ass:con}.\ref{item:contozero} is satisfied.
	
	At this point, notice that the consensus control law can be bounded as follows
	$\|k_{con}(\xi^{[i]}(k_0+n),\xi_{\mathcal{N}^{[i]}}(k_0+n))\|\leq \|L\xig \| = \|L \delta \|\leq \|L\|\|\delta\| = \|L\|\sqrt{\Phi(k_0+n)} \leq \|L\|((1-\lambda_{\Phi})^{n/2}\sqrt{\Phi(k_0)} + \sqrt{d}( e^{-\lambda^{\min}_\eta n} + (1-\lambda_{\Phi})^{n\over 2} )) \leq \|L\|((1-\lambda_{\Phi})^{n/2}\sum_{i\in\mathcal{I}}\|\xi_0^{[i]}\| + \sqrt{d}( e^{-\lambda^{\min}_\eta n} + (1-\lambda_{\Phi})^{n\over 2} ))=\beta(\sum_{i\in\mathcal{I}}\|\xi_0^{[i]}\|,n) $
	where the first equality comes from $L\one=0$, the third inequality is obtained by applying  twice the fact that $\sqrt{a+b}\leq \sqrt{a}+\sqrt{b}$ for any positive scalars $a,b>0$, and the fourth inequality stems from $\|\delta(k_0)\|\leq \|\xi(k_0)\|\leq  \sum_{i\in\mathcal{I}}\|\xi_0^{[i]}\|$. The function $\beta:\mathbb{R}_{\geq 0}\times \mathbb{R}_{\geq 0}\to \mathbb{R}_{\geq 0}$ is defined as
	$\beta(r,s):=\|L\|(1-\lambda_{\Phi})^{(\lfloor s\rfloor+k_0)/2}r + \|L\|\sqrt{d}( e^{-\lambda^{\min}_\eta (\lfloor s\rfloor+k_0)} + (1-\lambda_{\Phi})^{\lfloor s\rfloor+k_0\over 2})$, 
	$\forall s \in [\lfloor s\rfloor,\lfloor s\rfloor+1)$
	where $\lfloor s\rfloor$ denotes the smaller integer less than or equal to $s$, i.e., $\lfloor s\rfloor = \max_{k\in\mathbb{Z},\;k\leq s}k$. 
	 Since $\beta(r,s)$ is non increasing with $s$, the consensus control law is bounded as
	$\|k_{con}(\xi^{[i]}(k_0+n),\xi_{\mathcal{N}^{[i]}}(k_0+n))\|\leq \beta(\sum_{i\in\mathcal{I}}\|\xi_0^{[i]}\|,n)$ 
	uniformly over time, and  therefore Assumption~\ref{ass:con}.\ref{item:conbounded} is satisfied.
	Moreover, by definition of $\beta(\cdot)$ we have 
	$\int_{0}^{\infty}\beta(r,s) ds = \sum_{k=0}^{\infty}\beta(r,k) = 
	\sum_{k=k_0}^{\infty}\|L\|\left[(1-\lambda_{\Phi})^{ k/2}r + \sqrt{d}( e^{-\lambda^{\min}_\eta k}+ (1-\lambda_{\Phi})^{k\over 2})\right] 
	< \infty$
	where the last inequality holds from the fact that the summation converges since $|1-\lambda_{\Phi}|<1$ and $\lambda_\eta^{\min}>0$. Then, Assumption~\ref{ass:con}.\ref{item:betaint} holds and this concludes the proof. \hfill $\blacksquare$
	\vspace{-0.2cm}

\bibliographystyle{IEEEtran}

\bibliography{IEEEabrv,mybib}	
\end{document}